\documentclass[12pt]{iopart}
\pdfoutput=1
\usepackage{iopams}

\expandafter\let\csname equation*\endcsname\relax

\expandafter\let\csname endequation*\endcsname\relax

\usepackage{amsmath}

\usepackage{graphicx}
\usepackage[colorlinks, linkcolor=blue, citecolor=blue, urlcolor=blue, breaklinks=true]{hyperref}

\usepackage{etoolbox}
\usepackage{lipsum}% just for the example

\makeatletter
\def\@mkboth#1#2{}
\newlength\appendixwidth
\preto\appendix{\addtocontents{toc}{\protect\patchl@section}}
\newcommand{\patchl@section}{%
  \settowidth{\appendixwidth}{\textbf{Appendix }}%
  \addtolength{\appendixwidth}{1.5em}%
  \patchcmd{\l@section}{1.5em}{\appendixwidth}{}{\ddt}%
}
\makeatother

\usepackage{tikz}
\usetikzlibrary{shadows}
\usetikzlibrary{calc,decorations.pathmorphing,shapes}

\renewcommand{\eref}[1]{Eq.~(\ref{#1})}
\newcommand{\erefs}[1]{Eqs.~(\ref{#1})}
\renewcommand{\fref}[1]{Fig.~\ref{#1}}
\renewcommand{\sref}[1]{Sec.~\ref{#1}}
\renewcommand{\tref}[1]{Table~\ref{#1}}

\begin{document}

\title{Tracer particle diffusion in a system with hardcore interacting particles}

\author{Simon Pigeon$^{1,2,3}$, Karl Fogelmark$^3$, Bo S\"oderberg$^3$, Gautam Mukhopadhyay$^{4,5}$ and Tobias Ambj\"ornsson$^3$}

\address{$^1$ Laboratoire Kastler Brossel, UPMC-Sorbonne Universit\'es,
CNRS, ENS-PSL Research University, Coll\`ege de France,
4 place Jussieu Case 74, F-75005 Paris, France.}
\address{$^2$ Centre for Theoretical Atomic, Molecular and Optical Physics, School of Mathematics and Physics, Queen's University Belfast, Belfast BT7\,1NN, United Kingdom.}
\address{$^3$ Department of Astronomy and Theoretical Physics, Lund
  University, SE-223 62 Lund, Sweden.}
\address{$^4$ MU-DAE Centre for Excellence in Basic Science,
Mumbai-400098, India.}
\address{$^5$ Physics Department, Indian
  Institute of Technology, Bombay, Powai, Mumbai 400076, India.}
\ead{simon.pigeon@lkb.upmc.fr}
\vspace{10pt}

\begin{abstract}
%Diffusive transport plays a crucial role in a large variety of systems in science and especially in physics, chemistry and biology.
  In this study, inspired by the work of K. Nakazato and K. Kitahara
  [\href{http://ptp.oxfordjournals.org/cgi/doi/10.1143/PTP.64.2261}{\emph{Prog. Theor. Phys.}
    {\bf 64}, 2261 (1980)}], we consider the theoretical problem of
  tracer particle diffusion in an environment of diffusing hardcore
  interacting crowder particles. The tracer particle has a different
  diffusion constant from the crowder particles. Based on a
  transformation of the generating function, we provide an exact
  formal expansion for the tracer particle probability density, valid
  for any lattice in the thermodynamic limit. By applying this formal
  solution to dynamics on regular Bravais lattices we provide a
  closed form approximation for the tracer particle diffusion
  constant which extends the Nakazato and Kitahara results to include also b.c.c. and f.c.c. lattices. Finally, we compare our analytical results to simulations
  in two and three dimensions.
\end{abstract}

\tableofcontents

%\maketitle

\section{Introduction}

\begin{figure}
\begin{center}
\begin{tikzpicture}[scale=0.7,baseline=0]
\foreach \c in {  (-0.5, 3.5), (1.5, 1.5), (1.5, -2.5), (-3.5, -0.5), (4.5, -2.5), (-3.5, -0.5), (1.5, -2.5), (-3.5, 0.5), (-3.5, -0.5), (4.5, 1.5), (-2.5, 3.5), (3.5, 0.5), (0.5, 1.5), (-2.5, -0.5), (1.5, 0.5), (2.5, -0.5), (1.5, -2.5), (-1.5, 1.5), (-2.5, 3.5), (-3.5, -0.5), (1.5, -2.5), (4.5, 3.5), (3.5, 3.5), (-0.5, -1.5), (-1.5, 0.5), (-1.5, 1.5), (3.5, 1.5), (1.5, -0.5) }{
\node[circle,shading=ball, ball color=gray!20!white, minimum size=0.5cm] at \c {} ;}
\draw[step=1cm,gray,very thin] (-4.5,-3.5) grid (5.5,4.5);
\node[circle, opacity=0.2, fill =gray!, minimum size=0.5cm] at (-2.5,1.5) {} ;
\node[circle, opacity=0.2, fill =gray!, minimum size=0.5cm] at (-2.6,1.5) {} ;
\node[circle, opacity=0.2, fill =gray!, minimum size=0.5cm] at (-2.7,1.5) {} ;
\node[circle, opacity=0.2, fill =gray!, minimum size=0.5cm] at (-2.8,1.5) {} ;
\node[circle, opacity=0.2, fill =gray!, minimum size=0.5cm] at (-2.9,1.5) {} ;
\node[circle, opacity=0.2, fill =gray!, minimum size=0.5cm] at (-3,1.5) {} ;
\node[circle,shading=ball, ball color=gray!20!white, minimum size=0.5cm] at (-3.1,1.5) {} ;
\node[circle, opacity=0.2, fill =black!, minimum size=0.5cm] at (0.5,0.5) {} ;
\node[circle, opacity=0.2, fill =black!, minimum size=0.5cm] at (0.5,0.4) {} ;
\node[circle, opacity=0.2, fill =black!, minimum size=0.5cm] at (0.5,0.3) {} ;
\node[circle, opacity=0.2, fill =black!, minimum size=0.5cm] at (0.5,0.2) {} ;
\node[circle, opacity=0.2, fill =black!, minimum size=0.5cm] at (0.5,0.1) {} ;
\node[circle, opacity=0.2, fill =black!, minimum size=0.5cm] at (0.5,0.) {} ;
\node[circle,shading=ball, ball color=black!100!white, minimum size=0.5cm] at (0.5,-0.1) {} ;
\node[circle, opacity=0.2, fill =gray!, minimum size=0.5cm] at (2.5,2.5) {} ;
\node[circle, opacity=0.2, fill =gray!, minimum size=0.5cm] at (2.6,2.5) {} ;
\node[circle, opacity=0.2, fill =gray!, minimum size=0.5cm] at (2.7,2.5) {} ;
\node[circle, opacity=0.2, fill =gray!, minimum size=0.5cm] at (2.8,2.5) {} ;
\node[circle, opacity=0.2, fill =gray!, minimum size=0.5cm] at (2.9,2.5) {} ;
\node[circle, opacity=0.2, fill =gray!, minimum size=0.5cm] at (3,2.5) {} ;
\node[circle,shading=ball, ball color=gray!20!white, minimum size=0.5cm] at (3.1,2.5) {} ;
\node[circle, opacity=0.2, fill =gray!, minimum size=0.5cm] at (0.5,-2.1) {} ;
\node[circle, opacity=0.2, fill =gray!, minimum size=0.5cm] at (0.5,-2.2) {} ;
\node[circle, opacity=0.2, fill =gray!, minimum size=0.5cm] at (0.5,-2.3) {} ;
\node[circle, opacity=0.2, fill =gray!, minimum size=0.5cm] at (0.5,-2.4) {} ;
\node[circle, opacity=0.2, fill =gray!, minimum size=0.5cm] at (0.5,-2.5) {} ;
\node[circle, opacity=0.2, fill =gray!, minimum size=0.5cm] at (0.5,-2.) {} ;
\node[circle,shading=ball, ball color=gray!20!white, minimum size=0.5cm] at (0.5,-1.9) {} ;
\node[circle, opacity=0.2, fill =gray!, minimum size=0.5cm] at (-0.5,2.1) {} ;
\node[circle, opacity=0.2, fill =gray!, minimum size=0.5cm] at (-0.5,2.2) {} ;
\node[circle, opacity=0.2, fill =gray!, minimum size=0.5cm] at (-0.5,2.3) {} ;
\node[circle, opacity=0.2, fill =gray!, minimum size=0.5cm] at (-0.5,2.4) {} ;
\node[circle, opacity=0.2, fill =gray!, minimum size=0.5cm] at (-0.5,2.5) {} ;
\node[circle, opacity=0.2, fill =gray!, minimum size=0.5cm] at (-0.5,2.) {} ;
\node[circle,shading=ball, ball color=gray!20!white, minimum size=0.5cm] at (-0.5,1.9) {} ;
\node[circle, opacity=0.2, fill =gray!, minimum size=0.5cm] at (1.9, -1.5)  {} ;
\node[circle, opacity=0.2, fill =gray!, minimum size=0.5cm] at (1.8, -1.5) {} ;
\node[circle, opacity=0.2, fill =gray!, minimum size=0.5cm] at (1.7, -1.5)  {} ;
\node[circle, opacity=0.2, fill =gray!, minimum size=0.5cm] at (1.5, -1.5) {} ;
\node[circle, opacity=0.2, fill =gray!, minimum size=0.5cm] at (1.6, -1.5)  {} ;
\node[circle, opacity=0.2, fill =gray!, minimum size=0.5cm] at (2, -1.5)  {} ;
\node[circle,shading=ball, ball color=gray!20!white, minimum size=0.5cm] at (2.1, -1.5)  {} ;
\node[circle, opacity=0.2, fill =gray!, minimum size=0.5cm] at (4.5, -1.5)  {} ;
\node[circle, opacity=0.2, fill =gray!, minimum size=0.5cm] at (4.4, -1.5) {} ;
\node[circle, opacity=0.2, fill =gray!, minimum size=0.5cm] at (4.3, -1.5)  {} ;
\node[circle, opacity=0.2, fill =gray!, minimum size=0.5cm] at (4.2, -1.5) {} ;
\node[circle, opacity=0.2, fill =gray!, minimum size=0.5cm] at (4.1, -1.5)  {} ;
\node[circle, opacity=0.2, fill =gray!, minimum size=0.5cm] at (4., -1.5)  {} ;
\node[circle,shading=ball, ball color=gray!20!white, minimum size=0.5cm] at (3.9, -1.5)  {} ;
\node[circle, opacity=0.2, fill =gray!, minimum size=0.5cm] at (-3.5, -2.5)  {} ;
\node[circle, opacity=0.2, fill =gray!, minimum size=0.5cm] at (-3.5, -2.6) {} ;
\node[circle, opacity=0.2, fill =gray!, minimum size=0.5cm] at (-3.5, -2.7)  {} ;
\node[circle, opacity=0.2, fill =gray!, minimum size=0.5cm] at (-3.5, -2.8) {} ;
\node[circle, opacity=0.2, fill =gray!, minimum size=0.5cm] at (-3.5, -2.9)  {} ;
\node[circle, opacity=0.2, fill =gray!, minimum size=0.5cm] at (-3.5, -3)  {} ;
\node[circle,shading=ball, ball color=gray!20!white, minimum size=0.5cm] at (-3.5, -3.1)  {} ;
\draw[->,line width = 1pt] (-2.5,2) -- (-3.5,2);
\node at (-3,2.2) {$r_C$} ;
\draw[->,line width = 1pt] (0.,0.5)  -- (0.,-0.5) ;
\node at (-0.3,0.1) {$r_T$} ;
\end{tikzpicture}
\caption{Schematic illustration of the dynamics of a tracer particle
  in crowded system.  The dynamics takes place on a lattice where a
  set of crowder (white) particles attempt to move with jump rate
  $r_C$, and a tracer particle (black) attempts to move with jump rate
  $r_T$. The particles are subject to hard core repulsion, i.e. they
  cannot occupy the same site.  We here investigate the asymptotic
  statistical properties of the tracer particle motion in such a
  scenario.  Note that we here use a two-dimensional square lattice
  for illustrational purposes only -- the results derived herein apply
  to regular Bravais lattices in arbitrary number of
  dimensions.} \label{1}
\end{center}
\end{figure}
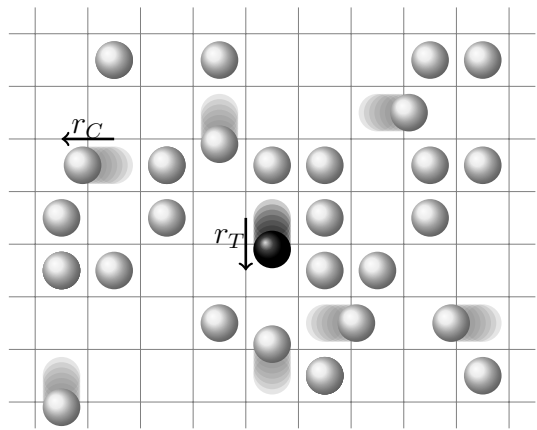

Diffusion of a particle in a crowded environment is a classical problem in statistical physics.
Among the different possible models, the simplest is to
consider the motion of a tracer particle on a lattice surrounded by
crowders at concentration $c$, see \fref{1}. In this model, which we explore herein,
 $n$ random walkers, each jumping on a lattice of $N$ sites in $d$ dimensions,
 surround a diffusing single (tagged) tracer particle. The tracer particle has a jump rate $r_T$
and each crowding particle has a jump rate $r_C$. Two particles cannot occupy
the same lattice site (hardcore repulsion).
It is intuitively clear from
\fref{1} that such crowding can severely change the effective diffusion
of a tracer particle as compared to a freely diffusing particle (no crowding particles).

In the mean field approximation the reduction
in the diffusion constant is obtained simply by counting the global average number of
available sites for the tracer particle; the diffusion constant then is
assumed to be reduced by a factor $1-c$, where $c = n/(N-1)$ is the
concentration of crowder particles. The mean field result is
a good approximation when $r_C\gg r_T$ in which case the tracer particle
at every instance in time ``experiences'' a
%BS: surrounding ->
local crowder concentration
equal to the global
average. For the general case, it is convenient to extract the mean
field result and write the tracer particle mean square displacement (MSD) for the crowded system as
\begin{equation}
\mathcal{MSD}(t) = 2d D_0t (1-c) f(t,c,r_T,r_C) \,, \label{Dc}
\end{equation}
where $D_0 = n_b a^2 r_T/(2d)$ corresponds to the diffusion constant of a single tracer
particle in the absence of crowder particles. Above, $n_b$ is the number of
nearest neighbours on the lattice and $a$ is the distance between such
neighbours. The quantity $f(t,c,r_T,r_C) $ is referred to as the correlation factor and it quantifies corrections to the mean field approximation. Thus, the correlation factor
contains all remaining information of the diffusion constant's dependence on the underlying physical
parameters, namely $r_T$, $r_C$ and $c$ and on the lattice geometry.
To illustrate the origin of the correlation factor, consider e.g.
that the tracer particle is
attempting to jump to a site occupied by a crowder particle. This move will be
prohibited due to the hardcore interaction. However, at the next time step the
tracer particle can attempt to jump to the same site. If the jump rate of the
crowder particle is such that $r_C/r_T\ll1$ it is probable that the same crowder
particle will inhibit this move once more. This type of two-body correlation effect is not
incorporated in the mean field approximation and, as we will show, in general,
leads to $f(t,c,r_T,r_C) < 1$.

Both the generality and complexity of the problem make its exact resolution
very challenging, if not impossible~\cite{haus1987}. Most of the
existing literature focus on specific cases. For example, if crowder
particles do not move ($r_C=0$) one has a single tagged particle
moving stochastically in a statically disordered medium
~\cite{havlin1987,Dean2004} (percolation).
Another well explored related scenario corresponds to single file
diffusion~\cite{bouchaud1990}, where in a linear lattice each particle
with identical jump rate ($r_C=r_T$) diffuse sub-linearly with
time. Moreover, diffusion of a forced tagged particle in a crowded
lattice is an actively explored problem
\cite{Burlatsky1992,Benichou1998,Illien2013,Benichou2013,Kundu2016}. Some
specific lattice structures were studied also, such as stripes and
capillaries \cite{Benichou2015}, presenting rich scaling behavior.

For the general case considered here (arbitrary concentration and
particle jump rates) only a few approximation
%BS added:
schemes
were
proposed~\cite{kikuchi1970,sankey1979,bender1979,fedders1977,nakazato1980,Brummelhuis1989}. More
significant results were obtained by Tahir-Kheli and
Elliott~\cite{tahir1983} based on a decoupling scheme of many-particle
correlation functions. The same results were expressed in terms of the
velocity autocorrelation function by van Beijeren and
Kutner~\cite{beijeren1985}. Some interesting results were more
recently obtained in the limit of high concentration
\cite{Benichou2000,Benichou2002} for forced tagged particle motion
\cite{Benichou2005,Benichou2014,Benichou2016}.

%Despite the importance of the problem considered here~\cite{haus1987}, only a
%few general analytical result based on a microscopic level were proposed. Most
%of the existing literature, focus on specific cases, for example where a single
%particle is moving stochastically in a statically disordered media
%($r_C=0$)~\cite{havlin1987,Dean2004} or focus on one-dimensional many-particle
%systems~\cite{bouchaud1990}. For the general case presented above
%(arbitrary concentrations and particles jump rates) only a few approximation
%were
%proposed~\cite{kikuchi1970,sankey1979,bender1979,fedders1977,nakazato1980}. More
%significant result were obtained by Tahir-Kheli and Elliott~\cite{tahir1983}
%based on a decoupling scheme of many-particle correlation functions. The same
%result were expressed in terms of the velocity autocorrelation function by van
%Beijeren and Kutner~\cite{beijeren1985}.

In our work, we provide all details of derivation of the transformed
(phase-rotated) Liouvillian operator introduced by Nakazato and
Kitahara \cite{nakazato1980}. To our knowledge, this full explicit
derivation has not been provided before. We then rigorously derive the
thermodynamic limit for the tracer particle probability density
function (PDF). Based on this PDF we obtain an exact expansion of the
correlation factor, $f(t,c,r_T,r_C)$, in terms of $n$-body correlation
functions, which is independent of the network or lattice
structure. By calculating the first order term in the formal expansion
we extend Nakazato and Kitahara's results \cite{nakazato1980} (valid
for linear, square and cubic lattices) to also include b.c.c. and f.c.c. lattices. Through extensive simulations we further
study in detail the limitation of our first-order approximation.  Our
simulation software package is made publically available. Since our
theoretical results are grounded in a formally exact result, it opens
up for future systematic improvements of our approximations.

The structure of this article is as follows: in \sref{def}, we
provide an exact expansion, in terms of many-body correlation
functions, for the probability density function (PDF) for the tracer
particle position in a crowded environment, valid for
arbitrary lattice geometries (\sref{definition}). We then present the
mathematical mapping used in this study (\sref{mapping}) and derive an
expression for the PDF in the thermodynamic limit (\sref{converge}). We
then discuss the dynamics in the mapped system (\sref{tiltdyna}) and
present a formal many-body expansion (\sref{expaa}).  The next \sref{regular} focuses on dynamics for regular lattices. For these
lattices, translation invariance allows a derivation of different
terms in the proposed expansion (\sref{mmssdd}). We then consider the
zeroth order of the expansion (\sref{zeross}), followed by the
determination of the crowding effect to the first order
(\sref{2body}). In the final \sref{simul}, we compare the
derived results to simulations for each of the lattices (in two and
three dimensions) considered.

\section{Many-body expansion for arbitrary lattices }\label{def}

In this section we will focus on the general mathematical formalism and
derivations for general lattice geometries. After some definitions using
the Dirac bra-ket notation (\sref{definition}) we map the system (\sref{mapping})
and derive an expression for the PDF in the thermodynamic limit
(\sref{converge}). It is followed by more details on the mapped system
dynamics (\sref{tiltdyna}) and our expansion for the PDF of the
tracer particle in terms of a sum of contributions from $n$ crowder particle
correlations (\sref{expaa}). The derivations in
Secs. \ref{mapping}-\ref{tiltdyna} follow closely those of Ref.~\cite{nakazato1980}, but are here provided in a slightly expanded format for
completeness.

\subsection{Hardcore interacting particles on a lattice}\label{definition}

We consider a simple lattice of $N$ lattice sites in $d$
dimensions. Each site is defined by a position vector
$\mathbf{u}$. Although we will use square lattices to illustrate the
system behavior and our mapping, what follows in this section is
independent of the lattice nature. The lattice contains one
tracer particle and $n$ crowder particles. A given microscopical state
of the system is thus fully characterized by the tracer particle
position, $\mathbf{r}$, and crowder particle positions, $\mathbf{r}_i$
with $i = \{1,\ldots,n\}$.  We use the bra-ket notation to define the
states and the dynamics of the system. For example, a state of the
system where only the tracer particle is present and located at
position $\mathbf{r}$ is denoted as:
\begin{equation}
  \vert T_{\mathbf{r}}\rangle \cdot \underbrace{\left( \prod_{\substack{ i = 1 \\ \mathbf{u}_i \neq \mathbf{r} }}^{N} \vert 0_{\mathbf{u_i}} \rangle \right)}_{\vert \{\emptyset\}\rangle}
\end{equation}
where $ \vert 0_{\mathbf{u_i}} \rangle $ is a local state
corresponding to no particle at site $\mathbf{u}_i$. A local site
state with a tracer particle present at position $\mathbf{r}$ is
denoted by $\vert T_{\mathbf{r}}\rangle $. A general allowed state
with both tracer and crowders present is written as the (direct)
product of all $N$ local states according to
%BS changed:
\begin{equation}
 \vert {T}_{\mathbf{r}} \rangle \cdot \left( \prod_{\substack{ i = 1
%    \\ \mathbf{r}_i \neq \mathbf{r}\\ \mathbf{r}_i \neq \mathbf{r}_j
}}^{n}
\vert C_{\mathbf{r}_i}\rangle \right) \cdot \vert \{\emptyset\}\rangle
\end{equation}
%BS inserted:
where it is understood that $\mathbf{r}$ and all $\mathbf{r}_i$ are
distinct, and
where $\vert \{\emptyset\}\rangle$ denotes remaining (non-occupied)
sites, i.e., the product of local states with neither tracer nor
crowder particles. Local crowder particle states are denoted by $\vert
C_{\mathbf{r}_i}\rangle$.  We use the following orthogonality rule
\cite{nakazato1980}
\begin{equation}
  \langle X_{\mathbf{r}} \vert \vert Y_{\mathbf{r}} \rangle = \delta_{X,Y}  \,, \label{ortho}
\end{equation}
where $X$ and $Y$ each represents one of the three types of lattice states
we consider: $T$ for tracer particle, $C$ for crowder particle or $0$
for an empty site. This orthogonality relation, together with our
explicit form for the Liouvillian below, handles the hard core
repulsion, forbidding two particles to occupy the same site on the
lattice. The completeness relation becomes:
\begin{equation}
I =\sum_{n =0}^{N-1} \sum_{\substack{\mathbf{r},\mathbf{r}_1,\ldots,\mathbf{r}_n\\ \mathbf{r}_i
    \neq \mathbf{r}\\ \mathbf{r}_i \neq \mathbf{r}_j}}  \vert {T}_{\mathbf{r}}
\rangle\vert C_{\mathbf{r}_1} \rangle \cdots
\vert C_{\mathbf{r}_n} \rangle \vert \{\emptyset\}\rangle \langle \{\emptyset\}\vert
\langle C_{\mathbf{r}_n}\vert \cdots  \langle C_{\mathbf{r}_1} \vert \langle {T}_{\mathbf{r}} \vert  \label{eq:completeness}
\end{equation}
where the sums run over all
%
%BS: Crowders indistinguishable? ->
% How many terms in sum? [binom(N-1,n) or (N-1)!/(N-n-1)!]
%
allowed crowder and tracer particle
configurations (crowder particles are indistinguishable). The quantity $I$ above is an identity operator which
when applied to any lattice state leaves that state unchanged. Notice,
however, the constraints in the sum above, which guarantee that
particles cannot occupy the same lattice site.

Let us now consider the dynamics. For a given particle type $P$ ($P$ = $T$ for
tracer or $C$ for crowder) we define an update operator according to
\begin{equation}
\mathcal{U}_P  = \sum_{i=1}^N   \sum_{\mathbf{b}} \Big( \vert
P_{\mathbf{u}_i+\mathbf{b}} \rangle \langle
0_{\mathbf{u}_i+\mathbf{b}} \vert \vert 0_{\mathbf{u}_i} \rangle
\langle P_{\mathbf{u}_i} \vert - \vert P_{\mathbf{u}_i} \rangle
\langle P_{\mathbf{u}_i} \vert \vert 0_{\mathbf{u}_i+\mathbf{b}}
\rangle \langle 0_{\mathbf{u}_i+\mathbf{b}} \vert \Big). \label{l}
\end{equation}
There are two sums above. The first sum is over all sites of the
lattice. The second sum is over all
neighboring sites of site $\mathbf{u}_i$,
%
%BS: r_i -> u_i?
%
where the quantity $\mathbf{b}$ is a displacement
vector to a nearest neighbor site. Within the
sums there are two terms. The first term corresponds to a scenario
where a particle of type $P$ on a site $\mathbf{u}_i$ is replaced by
an empty site ($P$ particle annihilated) followed by the creation of a
$P$ particle in $\mathbf{u}_i+\mathbf{b}$, if that site is empty. This
first term is thus a moving operator.  With the orthogonality relation
we defined in \eref{ortho} it is clear that the moving operator is
only effective if a $P$ particle is present in $\mathbf{u}_i$ and if
the site $\mathbf{u}_i+\mathbf{b}$ is empty, thus making sure that the
hardcore constraint is satisfied. The second term is negative
% BS changed:
% and produces no state change, and gives a non-zero
% result only when acting on a state
and contains the projection operator onto states with a $P$ particle
at $\mathbf{u}_i$ and an empty site at $\mathbf{u}_i+\mathbf{b}$.  These
two terms signify that in a small time window a particle can
either jump (first term) or remain at its current position (second
term). \fref{figL} schematically illustrates the superposition
of eight states produced by the action of the operator $\mathcal{U}_P$
on a square lattice containing a single $P$ particle.
\begin{figure}[h!]
\begin{center}
\begin{align}
\mathcal{U}_P \Big\vert \: \begin{tikzpicture}[scale=0.4,baseline=5]
  \node[circle,shading=ball, ball color=gray!20!white, minimum
    size=0.3cm] at (0.5,0.5) {}; \draw[step=1cm,gray] (-1.2,-1.2) grid
  (2.2,2.2);
\end{tikzpicture} \: \Big\rangle =  \Big\vert \: \begin{tikzpicture}[scale=0.4,baseline=5]
\node[circle,shading=ball, ball color=gray!20!white, minimum
  size=0.3cm] at (1.5,0.5) {}; \draw[step=1cm,gray] (-1.2,-1.2) grid
(2.2,2.2);
\end{tikzpicture} \: \Big\rangle + \Big\vert \: \begin{tikzpicture}[scale=0.4,baseline=5]
\node[circle,shading=ball, ball color=gray!20!white, minimum
  size=0.3cm] at (0.5,1.5) {}; \draw[step=1cm,gray] (-1.2,-1.2) grid
(2.2,2.2);
\end{tikzpicture} \: \Big\rangle + \Big\vert \: \begin{tikzpicture}[scale=0.4,baseline=5]
\node[circle,shading=ball, ball color=gray!20!white, minimum
  size=0.3cm] at (-0.5,0.5) {}; \draw[step=1cm,gray] (-1.2,-1.2) grid
(2.2,2.2);
\end{tikzpicture} \: \Big\rangle \nonumber \\
+ \Big\vert \:\begin{tikzpicture}[scale=0.4,baseline=5]
\node[circle,shading=ball, ball color=gray!20!white, minimum
  size=0.3cm] at (0.5,-0.5) {}; \draw[step=1cm,gray] (-1.2,-1.2) grid
(2.2,2.2);
\end{tikzpicture} \: \Big\rangle - 4 \Big\vert \:\begin{tikzpicture}[scale=0.4,baseline=5]
\node[circle,shading=ball, ball color=gray!20!white, minimum
  size=0.3cm] at (0.5,0.5) {}; \draw[step=1cm,gray] (-1.2,-1.2) grid
(2.2,2.2);
\end{tikzpicture} \: \Big\rangle \nonumber
\end{align}
\caption{Illustration of the action of the update operator
  $\mathcal{U}_P$, Eq.~\eqref{l}, applied to a square lattice
  with one particle $P$.} \label{figL}
\end{center}
\end{figure}

For the simplest case of a single tracer particle in the absence of
crowders, the probability density function (PDF) of finding the
particle in a position $\mathbf{r}$ starting from $\mathbf{r}_0$ at
time $t$ is
\begin{equation}\label{eq:PDF_single}
P(\mathbf{r},t\vert \mathbf{r}_0) = \langle \{
\emptyset \} \vert \langle {T}_{\mathbf{r}}\vert  e^{t r_T \mathcal{U}_T} \vert {T}_{\mathbf{r}_0}\rangle
\vert \{ \emptyset \}\rangle,
\end{equation}
where $r_P$ represents the jump rate of a $P$ particle for passing
to a nearest neighbour site. This is the
PDF for a single random walker on a lattice.

Our interest focuses on the diffusion of a tracer particle in a crowded
system with $n$ crowder particles. The PDF for finding the tracer
particle at position $\mathbf{r}$ after a time $t$, given an initial
position $\mathbf{r}_0$, then is:
  \begin{align}
P_{n,N}(\mathbf{r},t\vert \mathbf{r}_0)
&= \frac1{\binom{N-1}{n}}\langle \{\emptyset\}\vert  \langle {T}_{\mathbf{r}} \vert
\sum_{\mathbf{r}_1,\ldots,\mathbf{r}_n} \langle C_{\mathbf{r}_1}|\cdots
\langle C_{\mathbf{r}_n}\vert \nonumber\\
&\times e^{t\mathcal{L}}
\sum_{\mathbf{r}_1',\ldots,\mathbf{r}_n'} \vert C_{\mathbf{r}_1'}\rangle \cdots
\vert C_{\mathbf{r}_n'} \rangle \vert {T}_{\mathbf{r}_0}\rangle \vert \{\emptyset\}\rangle \label{pdf}\,
\end{align}
where $\sum_{\mathbf{r}_1,\ldots,\mathbf{r}_n}$ represents the sum
over all the possible configurations of the $n$ crowder particles on
the $N-1$ sites (thermal initial condition). The dynamics of our
system is determined by the Liouvillian operator
  $\mathcal{L} = r_C \mathcal{U}_C + r_T \mathcal{U}_T$, 
which allows either the crowder particles to move with a rate $r_C$
or the tracer particle to move with a rate $r_T$ at any given
instant. Due to the hardcore repulsion, care is required when
calculating \eref{pdf}. Formally, this expression can be evaluated by
dividing time into small time slices, i.e. we write
$\exp(t\mathcal{L})=\prod_{k=1}^K \mathcal{O}_k$ with
$\mathcal{O}_k=\exp[\epsilon_k\mathcal{L}]$ and $\epsilon_k =
t_k-t_{k-1}$, with $t_K=t$ and $t_0=0$. Between each
$\mathcal{O}_k$-operator in the product above, one then inserts
completeness relations, see \eref{eq:completeness}. Such an approach
(for $\epsilon_k\rightarrow 0$) ensures that at no time in $[0,t]$ do
any two particles occupy the same lattice site.  Due to the constraints
imposed in \eref{eq:completeness} such an approach is not practical,
however.  Instead, we here introduce a transformation which allows us
to rewrite the PDF above in terms of few-body states (to lowest orders
in an expansion). When the number of particles is one or two in the
transformed domain, we can deal with the hardcore repulsion by
introducing absorbing conditions for the many-body propagator.

\subsection{Mapping the system}\label{mapping}

Following Ref.~\cite{nakazato1980}
we now define the following generating function:
\begin{equation}
  G(\mathbf{r},t\vert \mathbf{r}_0;x) = \sum_{n=1}^{N} x^n
  P_{n,N}(\mathbf{r},t\vert \mathbf{r}_0) \binom{N-1}{n} \label{G}
\end{equation}
where $P_{n,N}(\mathbf{r},t\vert \mathbf{r}_0)$ is given in
\eref{pdf}.  The binomial coefficient
in the definition of the generating function corresponds to the number
of ways of placing the $n$ indistinguishable crowder particles on $N-1$ sites (one of
the $N$ sites is occupied by the tracer particle).  We will later find
it useful to normalize this generating function according to:
\begin{equation}
g(\mathbf{r},t\vert \mathbf{r}_0;x) =
\frac{1}{(1+x)^{N-1} } G(\mathbf{r},t\vert \mathbf{r}_0;x). \label{g}
\end{equation}

Using Eqs. (\ref{pdf}) and (\ref{G}) we have that \cite{nakazato1980}
\begin{equation}
  G(\mathbf{r},t\vert \mathbf{r}_0;x) =  \langle \{\emptyset\}\vert
  \langle {T}_{\mathbf{r}} \vert \prod_{\substack{ i = 1 \\
      \mathbf{u}_i \neq \mathbf{r} }}^{N} \Big(1 + \sqrt{x} \vert
  0_{\mathbf{u}_i} \rangle \langle
  C_{\mathbf{u}_i} \vert \Big) e^{t\mathcal{L}}  \prod_{\substack{ j = 1
    \\ \mathbf{u}_j \neq \mathbf{r}_0 }}^{N} \Big(1 + \sqrt{x} \vert
C_{\mathbf{u}_j} \rangle \langle 0_{\mathbf{u}_j} \vert \Big) \vert
{T}_{\mathbf{r}_0} \rangle \vert \{
\emptyset \} \rangle\,.\label{eq:sqrt_x_eq}
\end{equation}
The equivalence of \erefs{pdf} and (\ref{G}) to \eref{eq:sqrt_x_eq}
follows directly by explicitly writing out all terms in
\eref{eq:sqrt_x_eq}.  We require that $ \mathbf{u}_i \neq
\mathbf{r}$ and $ \mathbf{u}_j \neq
\mathbf{r}_0$, which above corresponds to the requirement that no crowder
particle can be placed on the same site as the tracer
particle. Following Ref.~\cite{nakazato1980}, we apply a transformation
(phase rotation) using the operator $\mathcal{S}_{\mathbf{u}_i} =
\vert C_{\mathbf{u}_i} \rangle \langle 0_{\mathbf{u}_i} \vert - \vert
0_{\mathbf{u}_i} \rangle \langle C_{\mathbf{u}_i} \vert $. Applying
$e^{-\theta\sum_{i=1}^{N}\mathcal{S}_{\mathbf{u}_i}}$, with a phase
parameter $\theta$, to the superposition of initial states (see \ref{sec:phase_rot}) we find that the generating
function~\eqref{g} can be conveniently written as
\begin{equation}\label{eq:g_Dirac}
  g(\mathbf{r},t\vert \mathbf{r}_0;x) = \langle \{\emptyset\} \vert \langle
  {T}_{\mathbf{r}}\vert e^{t\check{\mathcal{L}}(\theta)}
  \vert {T}_{\mathbf{r}_0} \rangle \vert \{ \emptyset \} \rangle \,
\end{equation}
provided we choose $\sqrt{x} = \tan \theta$. The transformed
Liouvillian is given by
\begin{equation}\label{eq:rot_L}
 \check{\mathcal{L}}(\theta)
= e^{-\theta \mathcal{S}} \mathcal{L} e^{\theta \mathcal{S}}.
\end{equation}
Thus, the phase rotation provides means for passing from a situation
with a large number of initial and final states (more precisely,
$\binom{N-1}{n}$ states, see \eref{pdf})
 to a situation with only one initial and one final
state. The final and initial states contain no crowder particles and
only one tracer particle. This reduction, however, comes at the
expense of a more complicated Liouvillian,
$\check{\mathcal{L}}(\theta)$.

\subsection{Thermodynamic limit}\label{converge}

We now seek to derive a general expression for the tracer particle's PDF in the
thermodynamic limit in terms of the phase-rotated generating function derived
in the previous subsection. To that end, let us write \erefs{G} and
\eqref{g} according to:
\begin{equation}
  g(\mathbf{r},t\vert \mathbf{r}_0;x) = \sum_{n=1}^{N} F_N(n,x) P_{c}(\mathbf{r},t\vert
  \mathbf{r}_0) \label{edece}\,,
\end{equation}
where
\begin{equation}
  F_N(n,x) = \frac{x^n}{(1+x)^{N-1} } \binom{N-1}{n} \,.
\end{equation}
In the limit of large $N$ we have that $ F_N(n,x)$ converges
to a normal distribution, i.e.,
\begin{equation}
  \lim_{N \rightarrow \infty} F_N(n,x) = \frac{1}{\sqrt{2\pi\sigma^2}}
  \exp\left(- \frac{(n-\tilde{n})^2}{2\sigma^2} \right) \,,
\end{equation}
with $\tilde{n} = N x/(1+x)$ and $\sigma^2 = N x/(1+x) $. 
Taking the thermodynamic
limit, i.e. taking $n$ to infinity with the concentration $c=n/N$ kept
fixed, we have:
\begin{equation}
  \lim_{\substack{
      N \rightarrow \infty  \\
      n \rightarrow \infty  \\
      \text{with } n/N = c } } F_N(n,x) = \frac1N \delta \left( c -
    \frac{  x}{(1+x)} \right) \,.
\end{equation}
Consequently, in the thermodynamic limit we choose
\begin{equation}
  x = \frac{c}{1-c} \label{criter} \,.
\end{equation}
Thus, in the thermodynamic limit the PDF becomes
\begin{equation}\label{eq:therm_limit}
  p_c(\mathbf{r},t\vert \mathbf{r}_0)=\lim_{\substack{
      N \rightarrow \infty  \\
      n \rightarrow \infty  \\
      \text{with } n/N = c } }  P_{n,N}(\mathbf{r},t\vert \mathbf{r}_0) =
  g(\mathbf{r},t\vert \mathbf{r}_0;x= \frac{c}{1-c}) \,,
\end{equation}
where $g(\mathbf{r},t\vert \mathbf{r}_0;x)$ is given in \eref{g}.  We
consequently have equality between the thermodynamic limit PDF of a
tracer particle moving through an environment of crowder
particles (undergoing usual displacement and hard core repulsion:
$\mathcal{L}$) and the PDF of a tracer particle moving without crowder
particles around and described by the modified Liouvillian
$\check{\mathcal{L}}(\theta)$. Using \eref{eq:g_Dirac} we write
\eref{eq:therm_limit} on the final form
\begin{equation}\label{eq:therm_limit2}
  p_{c}(\mathbf{r},t\vert \mathbf{r}_0) = \langle \{\emptyset\} \vert
  \langle  T_{\mathbf{r}} \vert e^{t\check{\mathcal{L}}(
    \theta)} \vert T_{\mathbf{r}_0} \rangle \vert \{ \emptyset \} \rangle
\end{equation}
where the phase angle is determined by $\theta = {\rm arctan}
[\sqrt{x}]= {\rm arctan} [\sqrt{c/(1-c)}]$.  Equation
\eref{eq:therm_limit2} gives the probability to find the tracer
particle at $\mathbf{r}$ starting from $\mathbf{r}_0$, after a time
$t$, in the thermodynamic limit knowing it is surrounded by a given
concentration $c$ of crowder particle (thermal initial conditions).
Notice that the result above is independent of the geometry of the
lattice, making it a powerful method to unravel crowding effect in
complex lattices and networks as well as in regular lattices. The
transformation is similar to a Lang--Firsov (polaron) transformation
used for electronic system \cite{Lang1963,mahan1993}.

\subsection{Phase-rotated dynamics of $\check{\mathcal{L}}$}\label{tiltdyna}

Let us now investigate further the dynamics described by the Liouvillian
$\check{\mathcal{L}}(\theta)$, see equation \eref{eq:rot_L}. This operator can
be written (see \ref{sec:phase_rot_L} and Ref.~\cite{nakazato1980})
\begin{equation}
\check{\mathcal{L}}(\theta) = r_C \mathcal{U}_C + r_T \cos^2\theta
\mathcal{U}_T + r_T \sin^2\theta \mathcal{R} + r_T \cos\theta \sin\theta
\left( \mathcal{C} + \mathcal{C}^\dagger\right) \label{ltilde}\,,
\end{equation}
where, in the thermodynamic limit, we have
$\cos^2 \theta \rightarrow (1-c)$, $\sin^2 \theta  \rightarrow c$ and $\cos \theta \sin
\theta \rightarrow \sqrt{c(1-c)}$ (because we have $\sqrt{x}=\tan
\theta$). The two first operators above were already formally defined through \eref{l}. The operator $\mathcal{R}$ acts by exchanging the position of
neighbor tracer and crowder particles:
\begin{equation}
\mathcal{R}  = \sum_{i=0}^N \sum_{\mathbf{b}}\Big( \vert
{T}_{\mathbf{u}_i} \rangle \langle C_{\mathbf{u}_i} \vert \vert
C_{\mathbf{u}_i+\mathbf{b}} \rangle \langle
{T}_{\mathbf{u}_i+\mathbf{b}} \vert - \vert
{T}_{\mathbf{u}_i} \rangle \langle {T}_{\mathbf{u}_i} \vert
\vert C_{\mathbf{u}_i+\mathbf{b}} \rangle \langle
C_{\mathbf{u}_i+\mathbf{b}} \vert \Big) \label{r}\,.
\end{equation}
Finally, the non-conservative operators $\mathcal{C}$ and $\mathcal{C}^\dagger$ are more
``exotic'' in the sense that they do not conserve the number of crowder
particles. These operators annihilate or create a crowder particle close to a
tracer particle, respectively.
\begin{align}
\mathcal{C}^\dagger &= \sum_{i=0}^N \sum_{\mathbf{b}}\Big( \vert
{T}_{\mathbf{u}_i} \rangle \langle {T}_{\mathbf{u}_i} \vert
\vert C_{\mathbf{u}_i+\mathbf{b}} \rangle \langle
0_{\mathbf{u}_i+\mathbf{b}} \vert - \vert {T}_{\mathbf{u}_i}
\rangle \langle 0_{\mathbf{u}_i} \vert \vert
C_{\mathbf{u}_i+\mathbf{b}} \rangle \langle
{T}_{\mathbf{u}_i+\mathbf{b}} \vert \Big) \label{t}\,,
\\ \mathcal{C} &= \sum_{i=0}^N \sum_{\mathbf{b}}\Big( \vert
{T}_{\mathbf{u}_i} \rangle \langle {T}_{\mathbf{u}_i} \vert
\vert 0_{\mathbf{u}_i+\mathbf{b}} \rangle \langle
C_{\mathbf{u}_i+\mathbf{b}} \vert - \vert {T}_{\mathbf{u}_i}
\rangle \langle C_{\mathbf{u}_i} \vert \vert
0_{\mathbf{u}_i+\mathbf{b}} \rangle \langle
{T}_{\mathbf{u}_i+\mathbf{b}} \vert \Big) \label{t+}\,.
\end{align}
Here the $\mathcal{C}^\dagger$ operator takes effect only when applied
to states with a tracer particle close to an empty site (like
$\mathcal{U}_T$), while the operator $\mathcal{C}$ takes effect only
when applied to states with a crowder particle close to the tracer
particle (like $\mathcal{R}$).

Thus, the dynamics in phase-rotated space as described by
$\check{\mathcal{L}}(\theta)$ is very different from the
dynamics described by $\mathcal{L}$. Despite having reduced the
thermal initial states to single particle states, we have, with
$\check{\mathcal{L}}(\theta)$, introduced the possibility for the
tracer particle to swap positions with a crowder (through
$\mathcal{R}$), and to create and annihilate crowder particles in the
neighborhood of the tracer particle (through the operators
$\mathcal{C}^\dagger$ and $\mathcal{C}$).

\subsection{Expansion in terms of particle correlation functions}\label{expaa}

In this subsection we present a formally exact expansion of the tracer
particle PDF. However, before giving the formal expansion, let us discuss the
underlying ``philosophy'' of the expansion. For a tracer particle surrounded
by other particles, at ``zeroth order'' the most relevant effect will be to
determine if the next site to jump to is free or not. This effect is
contained in the mean field approximation, as previously described. Our
expansion is constructed to give the mean field to lowest order. At the next
order, as discussed in the introduction, we may find a situation where the
tracer particle attempts to jump consecutively to a site where some crowder
particle is present. Including this effect, goes beyond mean field, and is
contained to next order in our expansion. Higher order terms in our expansion
include more complex effects. Such effects are especially prominent for
comparatively slow crowder particles, $r=r_C/r_T\ll1$ 
 ($r\rightarrow 0$ is the percolation limit).

Let us now introduce our formal expansion. To that end we take the
Laplace transform with respect to time of ${p}_{c}(\mathbf{r},t\vert \mathbf{r}_0)$ in
\eref{eq:therm_limit2}:
\begin{equation}
 \tilde{p}_{c}(\mathbf{r},s\vert \mathbf{r}_0) = \langle \{\emptyset\} \vert \langle
  T_{\mathbf{r}} \vert \frac1{s-\check{\mathcal{L}}} \vert
  T_{\mathbf{r}_0}\rangle
  \vert \{ \emptyset \} \rangle \label{pdfz2}\,.
\end{equation}
We proceed by splitting the phase-rotated Liouvillian according to
$\check{\mathcal{L}} = \check{\mathcal{L}}_0 +
\gamma\check{\mathcal{L}}'$, where $\check{\mathcal{L}}_0 = r_C
\mathcal{U}_C + r_T(1-c) \mathcal{U}_T +r_T c \mathcal{R}$ and
$\check{\mathcal{L}}'= \mathcal{C}+ \mathcal{C}^\dagger$ correspond to
the conservative part and the non-conservative part with respect to
the number of crowder particles, respectively, with $\gamma=
r_T\sqrt{c(1-c)}$. By repeatedly using the identity:
\begin{equation}
  \frac1{s-\check{\mathcal{L}}} = \frac1{s-\check{\mathcal{L}}_0} +\gamma \frac1{s-\check{\mathcal{L}}_0} \check{\mathcal{L}}'\frac1{s-\check{\mathcal{L}}}\,,
\end{equation}
we can write our formal expansion
\begin{equation}
  \tilde{p}_c(\mathbf{r},s\vert \mathbf{r}_0) = \sum_{j=0}^\infty
  \gamma^{2j}\tilde{\phi}_{j}(\mathbf{r},s\vert \mathbf{r}_0),
  \label{pexpansion}
\end{equation}
where
\begin{equation}
  \tilde{\phi}_{j}(\mathbf{r},s\vert \mathbf{r}_0) = \langle
  \{ \emptyset \}\vert\langle {T}_\mathbf{r} \vert\frac1{s -
    \check{\mathcal{L}}_0}  \left( \check{\mathcal{L}}'
    \frac1{s-\check{\mathcal{L}}_0} \right)^{2j} \vert {T}_{\mathbf{r}_0}\rangle 
  \vert \{ \emptyset \} \rangle.
  \label{expand}
\end{equation}
The absence of odd powers
% of $j$
of $\gamma$ in the sum in \eref{pexpansion} appears because the
$\check{\mathcal{L}}'$ operator creates or destructs only one crowder
particle at each application. As neither the initial nor the final
states contain crowder particles, we need an even number of
applications of this operator to yield a non-zero contribution.  We
notice that the expansion parameter,
\begin{align}\label{eq:gamma2} \gamma^2= r_T^2 c(1-c)\le r_T^2/4,
\end{align}
converges to zero for small ($c\rightarrow 0$) or large ($c\rightarrow
1$) crowder concentrations. However, notice (see
Eqs. (\ref{ltilde}), (\ref{pexpansion}) and (\ref{expand})) that the
quantity $\tilde{\phi}_{j}(\mathbf{r},s\vert \mathbf{r}_0)$ also (in a
non-trivial fashion) depends on $r$ and $c$. Therefore, care is
required when studying the $c\rightarrow 0$ and $ c\rightarrow 1$
limits.

To clarify the underlying physics of each term in \eref{pexpansion}
let us consider the first few ones. For $j=0$ we have, in the time domain,
\begin{equation} {\phi}_{0}(\mathbf{r},t\vert \mathbf{r}_0) = \langle
  \{ \emptyset \}\vert\langle {T}_\mathbf{r} \vert e^{t(1-c) r_T
    \mathcal{U}_T } \vert {T}_{\mathbf{r}_0}\rangle \vert \{ \emptyset
  \}\rangle \label{expand0}\,.
\end{equation}
This expression is the PDF for diffusion of the tracer particle where
the jump rate is renormalized by the concentration of empty sites
($1-c$), compared to the single particle PDF in
\eref{eq:PDF_single}. This thus corresponds to the mean field
approximation discussed in the introduction. The fact that only the
tracer particle update operator, $r_T \mathcal{U}_T$, enters the
expression above, and not the full $\check{\mathcal{L}}_0$, follows
from the fact that the operators $\mathcal{U}_C$ and $\mathcal{R}$ act
on crowder particle states and that the phase-rotated initial state
does not include crowder particles.  Note that the thermodynamic
result above is independent of the lattice considered.

Going further to the first order correction ($j=1$) we have the
Laplace domain result
\begin{equation}
  \tilde{\phi}_{1}(\mathbf{r},s\vert \mathbf{r}_0) =  \langle \{
  \emptyset \}\vert\langle {T}_\mathbf{r} \vert \frac{1}{s-
    r_T(1-c)\mathcal{U}_T } \mathcal{C}
  \frac{1}{s-\check{\mathcal{L}}_0}  \mathcal{C}^\dagger \frac{1}{s- r_T(1-c)\mathcal{U}_T }\vert
  {T}_{\mathbf{r}_0}\rangle \vert \{ \emptyset
  \}\rangle. \label{expand2}
\end{equation}
The propagator above corresponds to a process where a tracer particle
diffuses with a renormalized jump rate (mean field); at a given time a
crowder particle is created followed by the evolution of both
particles according to $\check{\mathcal{L}}_0$. Thereafter the crowder
particle is annihilated, followed by the diffusion of the tracer
particle. 
%
%\begin{figure}
%\begin{center}
%\includegraphics[width=0.4\textwidth]{diagram.pdf}
%\caption{Diagrammatic representation of the few body correlation contribution
%  $\tilde{\phi}_{2}$
%  where two crowder particle are created. The vertical segment correspond to evolution according to $\check{\mathcal{L}}_0$. In (a.) the two crowder particle are consecutively created before being annihilated, while in (b.) they are successively created and annihilated. Notice in diagram (a.) in the middle the system count two indistinguishable crowder particles.}\label{diag}
%\end{center}
%\end{figure}

Considering higher order terms, $j\ge 2$ in \eref{pexpansion}, we
notice that the reduction we made for the operator
$\check{\mathcal{L}}'$ for $\tilde{\phi}_{1}$ now is more complicated.
For example, for $j=2$ we find that $\tilde{\phi}_{2}$ can be
decomposed into two distinct processes:
(i) one involving the consecutive creation of two crowder particles
followed by their annihilation, and
(ii) another one where a crowder particle is created and then
destroyed, followed by the creation and subsequent annihilation of a
new crowder. This corresponds to the case where the tracer particle
consecutively meets two particles. When evaluating higher order terms
(arbitrary $j$) one must make sure that the creation and annihilation
operators originating from $\check{\mathcal{L}}'$ are constrained such
that no more crowders are annihilated than created. Note that all
combinations are embedded into the probability $\tilde{\phi}_{j}$:
(i) creating $j$ crowder particles and then annihilating them,
(ii) the process of creation of a crowder directly followed by its
annihilation, $j$ times, and
(iii) all intermediate combinations. 
The expansion proposed in \eref{pexpansion} is based on the
  assumption that $\gamma$ is ``small''. In practice, the simulation results in
  \sref{simul}  indicate that the convergence of \eref{pexpansion} may be
  controlled  by $\gamma/r_C$ (for smaller values of this ratio, the first-order approximation becomes better).

It is worth noticing that if the zeroth and first order terms
  relate to diffusion of one and two interacting particles,
  respectively, already the second order, involving the diffusion of
  three interacting particles, is beyond reach of exact calculation in
  general.  In the related context of diffusion in disordered media,
a somewhat similar diagrammatic expansion was studied using renormalization
group methods \cite{Dean2004}.

In the next \sref{regular}, we will see that for a regular lattice,
analysis of the exact expansion provides good estimates for the tracer
particle diffusion coefficient.

\section{Crowding effects on tracer particle diffusion on regular lattices}\label{regular}

In this section we focus on diffusion on regular lattices where the
separation to nearest neighbors is homogeneous. These lattices fall into
the Bravais lattice category, and have translation
invariance. In \tref{lambda} we present the definition of the
different neighbor displacements, $\mathbf{b}$. In \tref{tabla} other
characteristic numbers defining the lattice geometry are
provided. Below, we first relate the PDF introduced previously to the
tracer particle MSD (\sref{mmssdd}). We
then derive the zeroth order contribution to the diffusion constant in
\sref{zeross}, followed by results for the first order contribution
(\sref{2body}).
\begin{table}
\begin{center}
 
 \begin{tabular}{|c|c|c|c|}
\hline
Lattice & $n_b$ & Vector connecting the nearest neighbors is $\mathbf{b}=\pm a\mathbf{v}$ \\ \hline\hline
Linear & $2$&  $\mathbf{v}=1$ \\ \hline\hline
Square & $4$&  $  \mathbf{v}=\left\{ \begin{pmatrix} 1 \\ 0 \end{pmatrix}, \begin{pmatrix} 0 \\ 1 \end{pmatrix}\right\}$  \\ \hline
%Hexagonal & $6$&   $ \left\{  \begin{pmatrix} 0 \\ \pm 1 \end{pmatrix},  \begin{pmatrix} \pm \sqrt{3}/2 \\ 1/2 \end{pmatrix}, \begin{pmatrix} \pm \sqrt{3}/2 \\ -1/2 \end{pmatrix} \right\}$  \\ \hline 
\hline
Cubic & $6$&  $ \mathbf{v}=\left\{ \begin{pmatrix} 1 \\ 0 \\ 0 \end{pmatrix}, \begin{pmatrix} 0 \\ 1 \\ 0 \end{pmatrix}, \begin{pmatrix} 0 \\ 0 \\ 1 \end{pmatrix} \right\} $  \\ \hline
B.c.c. & $8$&  $ \mathbf{v}=\dfrac{1}{\sqrt{3}}  \left\{ \begin{pmatrix} 1 \\ 1 \\ 1 \end{pmatrix},   \begin{pmatrix} - 1 \\ 1 \\ 1 \end{pmatrix},  \begin{pmatrix}  1 \\ -1 \\ 1 \end{pmatrix}, \begin{pmatrix} 1 \\ 1 \\ -1 \end{pmatrix} \right\} $  \\ \hline
F.c.c. & $12$&  $ \mathbf{v}=\dfrac{1}{\sqrt{2}}\left\{ \begin{pmatrix}  1 \\ 1 \\ 0 \end{pmatrix}, \begin{pmatrix} -1 \\  1 \\ 0 \end{pmatrix}, \begin{pmatrix} 1 \\ 0 \\ 1 \end{pmatrix},  \begin{pmatrix}  -1 \\ 0 \\ 1 \end{pmatrix}, \begin{pmatrix} 0 \\ 1 \\ 1 \end{pmatrix}, \begin{pmatrix} 0 \\ -1 \\ 1 \end{pmatrix}   \right\} $  \\ \hline
\end{tabular}
\caption{List of the nearest neighbor displacement vector, $\mathbf{b}$, and
  the number of nearest neighbours,  $n_b$, for the different considered  lattices, where $a$ is the distance between nearest neighbours. }\label{lambda}
\end{center}
\end{table}

\subsection{Mean square displacement on regular lattices}\label{mmssdd}

For regular lattices we are now interested in the MSD of a tracer
particle. Using the definition of the variance of $p_c(\mathbf{r},t\vert
\mathbf{r}_0)$ and the expansion in \eref{pexpansion} we obtain
\begin{equation}
  \mathcal{MSD}(t)=\sum_{j=0}^\infty  \gamma^{2j}   A_{j}(t)t,
\label{dfffd}
\end{equation}
where
\begin{equation}
  A_{j}(t) = \frac1t \sum_{\mathbf{n},\mathbf{n}_0} (\mathbf{r}(t)-\mathbf{r}_0)^2
  \phi_{j}(\mathbf{r},t\vert \mathbf{r}_0).
\label{aj}
\end{equation}
Using the definition of the correlation factor, \eref{Dc}, we have
\begin{equation}
  f(t,c,r_C,r_T) =  \tfrac{1}{n_b r_T a^2(1-c)}\sum_{j=0}^\infty  \gamma^{2j} A_{j}(t).
\label{dcdcd}
\end{equation}
To proceed, we introduce the Fourier transform of the position of
the tracer particle:
\begin{equation}
\vert \hat{T}_\mathbf{q} \rangle =\sum_{\mathbf{r}} e^{-i\mathbf{q\cdot r}} \vert {T}_\mathbf{r}\rangle 
\end{equation}
with inverse
\begin{equation}
   \vert {T}_\mathbf{r} \rangle = \frac{1}{(2\pi)^d} \int d^dq  e^{i\mathbf{q \cdot r}}\vert \hat{T}_\mathbf{q} \rangle ,
\end{equation}
where $\int d^dq=(l_1\cdots l_d) \int_{-\pi/l_1}^{\pi/l_1}dq_1\cdots
\int_{-\pi/l_d}^{\pi/l_d}dq_d$ with $l_i = a v_i$ ($i=1,\ldots,d$) and $v_i$
the $i$th component of neighboring vector $\mathbf{v}$ as given in \tref{lambda}.
The Fourier transform in space associated with
$p_c(\mathbf{r},t\vert \mathbf{r}_0)$ is
\begin{align}
  S(\mathbf{q},t) = \frac{1}{N} \sum_{\mathbf{r},\mathbf{r}_0}
  e^{-i\mathbf{q\cdot}(\mathbf{r}-\mathbf{r}_0)} \langle \{ \emptyset
  \}\vert\langle {T}_\mathbf{r} \vert e^{t\check{\mathcal{L}}} \vert
  {T}_{\mathbf{r}_0}\rangle \vert \{ \emptyset \} \rangle.
\end{align}
Using the expansion in \eref{pexpansion} of $p_c(\mathbf{r},t\vert
\mathbf{r}_0)$ we have that
\begin{align} {S}(\mathbf{q},t) = \sum_{j=0}^\infty \gamma^{2j}
  {S}_{j}(\mathbf{q},t),
\label{Sqt}
\end{align}
where
\begin{equation}
  S_j(\mathbf{q},t) =  \frac{1}{N} \sum_{\mathbf{r},\mathbf{r}_0}
  e^{-i\mathbf{q\cdot}(\mathbf{r}-\mathbf{r}_0)}
  \tilde{\phi}_{j}(\mathbf{r},t\vert \mathbf{r}_0)\label{Siqt}
\end{equation}
are the Fourier transforms related to each
$\tilde{\phi}_{j}(\mathbf{r},s\vert \mathbf{r}_0)$ as defined
earlier. The MSD now can be extracted from ${S}(\mathbf{q},t)$ as the
second order term in $\mathbf{q}$, according to
\begin{equation}
  S(\mathbf{q},t) =  1 - \frac{|\mathbf{q}|^2}{2d} \mathcal{MSD}(t)
  + O[\mathbf{q}^3]
\label{msdt}
\end{equation}
where $|\mathbf{q}|^2 = q_x^2 + q_y^2 + q_z^2 $. The fact that no cross-terms ($q_xq_y$ etc) appear above follows from the fact that the dynamics for all lattices considered herein is rotationally symmetric. The contribution to the crowding effect then is
contained in
\begin{equation}
  \bar{A}_{j} = \lim_{t\to\infty} A_{j}(t) =- \lim_{t\to\infty}\frac1{t}
  \nabla^2_{\mathbf{q}}S_j(\mathbf{q},t) \Big\vert_{\mathbf{0}}, \label{chat}
\end{equation}
with $S_j(q,t)$ given by \erefs{expand} and \eqref{Siqt}.

\subsection{Zeroth order: mean field approximation}\label{zeross}

The zeroth order term in the expansion in the previous subsection is
straightforward to derive, because no crowder particle states are
involved. The Laplace transform is (see \erefs{expand0} and \eqref{Siqt})
  \begin{align}
\tilde{S}_0(\mathbf{q},s) &= \frac{1}{N} \sum_{\mathbf{r},\mathbf{r}_0}
e^{-i\mathbf{q\cdot}(\mathbf{r}-\mathbf{r}_0)} \langle \{ \emptyset \}\vert\langle {T}_\mathbf{r}
\vert \frac1{s-r_T(1-c)\mathcal{U}_T} \vert {T}_{\mathbf{r}_0}\rangle \vert \{
\emptyset \} \rangle \nonumber\\
& = \frac{1}{s-  r_T(1-c) \sum_{\mathbf{b}} (e^{i \mathbf{q\cdot b}} - 1) }\,,
\end{align}
with $\sum_{\mathbf{b}} $ the sum over all the neighbor displacements of
the considered lattice.  We have that $\sum_{\mathbf{b}} \left( e^{i\mathbf{q\cdot b}} -1
\right)\approx -a^2 \sum_{\mathbf{v}}(\mathbf{q}\cdot\mathbf{v})^2 = -n_b a^2 |\mathbf{q}|^2/(2d)$, where in the last step we used the explicit expressions for $\mathbf{v}$ in \tref{lambda}. We thus obtain
\begin{equation}\label{eq:A0}
A_{0}(t)=A_0= a^2 n_b r_T(1-c)\,,
 \end{equation}
where $n_b$ is the number of nearest neighbors.
Thus, the lowest order term in our expansion is indeed the mean field result,
where the crowded environment provides a renormalization of the diffusion constant for
the tracer particle by a factor $(1-c)$.  As stated previously, the mean field
result is a useful approximation as long as $\gamma^2=r_T^2c(1-c)$ is
``small''.

Let us now venture beyond the mean field approximation, into the first order of the
expansion.

\subsection{First order: including two-body correlation effects}\label{2body}

Consider now the first order
%
%BS changed (removed "based" (correct?) and inserted commas):
%term based $p_1(\mathbf{r},t\vert\mathbf{r}_0 )$
term, $p_1(\mathbf{r},t\vert\mathbf{r}_0 )$,
as given in \erefs{expand2} and \eqref{Siqt}. We have for the
Laplace-Fourier transform that
\begin{equation}
  \tilde{S}_1(\mathbf{q},s) =  [\tilde{S}_0(\mathbf{q},s)]^2  \frac{1}{N}\sum_{\mathbf{r},\mathbf{r}_0}
  e^{-i\mathbf{q\cdot}(\mathbf{r}-\mathbf{r}_0)} \langle \{ \emptyset
  \}\vert\langle {T}_\mathbf{r} \vert \mathcal{C}
  \frac1{s-\check{\mathcal{L}}_0} \mathcal{C}^\dagger \vert
  {T}_{\mathbf{r}_0}\rangle \vert \{ \emptyset
  \}\rangle.
\label{s1}
\end{equation}
As shown above, we have that $\tilde{S}_0^2(\mathbf{q},s)\sim
1/s^2+\mathcal{O}[\mathbf{q}^2]$ leading to
\begin{equation}
  A_{1}(t)t   =  {L}^{-1}\left\{ \frac{1}{s^2}\frac{-4n_b a^2 V(s)}
    {1-2(r_T(1-3c) + r_C ) V(s)  } \right\},
\label{A1}
\end{equation}
as derived in \ref{aa}. $V(s)$ depends on the lattice and is defined
respectively for the hypercubic (linear, square and cubic), b.c.c. and
f.c.c. lattices according to:
\begin{align}
V_{\text{hyp}}(s) & = \int_{-\pi}^{\pi} \frac{dp_x}{2\pi} \int_{-\pi}^{\pi} \frac{dp_y}{2\pi}\int_{-\pi}^{\pi} \frac{dp_z}{2\pi}
\frac{\sin^{2}(p_{x})}{s-2(r_{T}(1-c)+r_{C})\sum_{\mathbf{v}}(\cos(\mathbf{p}\cdot\mathbf{v})-1)} \label{cub}\\
V_{\text{bcc}}(s) & =4\int_{-\pi}^{\pi} \frac{dp_x}{2\pi} \int_{-\pi}^{\pi} \frac{dp_y}{2\pi}\int_{-\pi}^{\pi} \frac{dp_z}{2\pi}\frac{\sin^{2}(p_{x}/\sqrt{3})\cos^{2}(p_{y}/\sqrt{3})\cos^{2}(p_{z}/\sqrt{3})}{s-2(r_{T}(1-c)+r_{C})\sum_{\mathbf{v}}(\cos(\mathbf{p}\cdot \mathbf{v})-1)} \label{bcc}\\
V_{\text{fcc}}(s) & =2\int_{-\pi}^{\pi} \frac{dp_x}{2\pi} \int_{-\pi}^{\pi} \frac{dp_y}{2\pi}\int_{-\pi}^{\pi} \frac{dp_z}{2\pi}\frac{\sin^{2}(p_{x}/\sqrt{2})\cos(p_{y}/\sqrt{2})\left(\cos(p_{y}/\sqrt{2})+\cos(p_{z}/\sqrt{2})\right)}{s-2(r_{T}(1-c)+r_{C})\sum_{\mathbf{v}}(\cos(\mathbf{p}\cdot \mathbf{v})-1)}\label{fcc}
\end{align}
with $r=r_C/r_T$ is the relative crowder particle jump rate and where the unit vectors $\mathbf{v}$ are listed in \tref{lambda} for the different lattice types considered. For linear and square lattices the integrals in $V_{\text{hyp}}(s)$ are replaced by integrals over $p_x$ and over $(p_x,p_y)$, respectively. 
Similar results were derived for translationally non-invariant
lattices, such as honeycomb and diamond, in Ref. \cite{Suzuki2002}.
There, also some details were given regarding the derivation for
linear, square and cubic lattices. For these, the results given above
match those of Nakazato and Kitahara \cite{nakazato1980}. The
derivation provided here extends the Nakazato and Kitahara results
 to f.c.c. and b.c.c. lattices.

Taking into account only the zeroth and first order of the expansion
(\ref{dcdcd}) we find the first order approximation of the correlation
factor
\begin{align}
  f_{\text{first-order}}(t) &= 1 + \gamma^2
  \frac{A_{1}(t)}{A_0}\label{l1d0} \\&= 1 -
  \frac{1}{t}{L}^{-1}\left\{ \frac{1}{s^2}\frac{ 4 r_T c V(s)  }{1-
      2 (r_T(1-3c) + r_C ) V(s) } \right\}.
\label{l1d}
\end{align}
where we used the mean field result in \eref{eq:A0}, and $A_1(t)$ is
given in \eref{A1}. The results above generalize the Nakazato and Kitahara
results \cite{nakazato1980} to a larger class of lattices. 

We will later show by simulations (Sec. \ref{simul}) that the
first-order approximation given above overestimates the true
correlation factor for the cases considered. This finding is
consistent with (but does not prove) that higher order terms in the
expression (\ref{dcdcd}) (second order and beyond)
%
%BS: ???
linked to
high order correlation effect, contribute negatively to the
diffusion, i.e., that $A_j(t)<0$ also for $j\ge 2$.

%In appendix X, we use this assumption to also provide an estimated lower bound for the correlation factor. }

For a one dimensional lattice the coefficient $V(s)$,
\eref{cub}, can be estimated yielding a correlation factor (see \ref{tt})
\begin{equation}
  f_{\text{first-order}}(t)   \underset{t \rightarrow \infty  }{=}
  \frac{1}{c} \sqrt{\frac{ r+1-c }{ \pi r_T  t}}.
\label{flong}
\end{equation}
where $r=r_C/r_T$.
Existing exact results are $f(t \rightarrow \infty) = \tfrac{1}{c}
\sqrt{r_{\rm eff}/(r_T^2 \pi t)} $ asymptotically
\cite{Harris1965,Goncalves2008}, where the quantity $r_{\rm
  eff}=(n+1)r_Cr_T/( r_C+ nr_T)$. In the thermodynamic limit,
$n\rightarrow \infty$ as considered here, we hence have $r_{\rm
  eff}=r_C$. We thus notice that for this case indeed
$f(t)<f_{\text{first-order}}(t)$, and the scaling with time agrees
with exact results, i.e., $\mathcal{MSD}(t)\underset{t \rightarrow
  \infty }{\sim} t^{1/2}$.

For higher dimensions we are restricted to the long time
  limit which becomes (see \ref{tt}):
\begin{equation}
  f_{\text{first-order}}(\infty) = 1 - \frac{2c}{2c + (r + 1 - c)\beta},
\label{fmamm}
\end{equation}
with $\left(\beta+1\right)^{-1} = 2r_T(r+1-c)  V(0) $, where
$\beta$ is a structure factor depending on the lattice geometry, and
$r=r_C/r_T$ is the relative crowder particle jump rate as before.  The
values of the structure factors are listed in \tref{tabla} for
the considered lattices, together with other typical characteristics
(dimension and nearest neighbor count).
For the one-dimensional (linear) lattice we have
  $\beta=0$, giving $f_{\text{first-order}}(\infty) =0$,
  indicating the sub-linear scaling of the MSD with time as discussed
  previously. For the other lattices we have
  $f_{\text{first-order}}(\infty) \neq 0$,
  representing linear scaling of the MSD.
\begin{table}
\begin{center}
\begin{tabular}{|c|c|c|c|}
  \hline Lattice   & $\quad d \quad$ & $\quad n_b \quad$ & $\beta$ \\
  \hline
  \hline Linear    & 1               & 2               & 0 \\
  \hline
  \hline Square    & 2               & 4               & $2/(\pi-2)$ \\
%  \hline Hexagonal & 2               & 6               & $3.0253$ \\
  \hline
  \hline Cubic     & 3               & 6               & $3.7655$ \\
  \hline B.c.c.    & 3               & 8               & $5.04863$\\%$6.1028$ \\
  \hline F.c.c.    & 3               & 12              & $8.24308$\\%$9.63151$ \\
  \hline
\end{tabular}
\caption{Table of lattice characteristics, where $d$ is the
  dimension of the lattice, $n$ the nearest neighbor count, and
  $\beta$ a structural factor appearing in the correlation
  factor, Eq.~\eqref{fmamm}.}
\label{tabla}
\end{center}
\end{table}
In the next section, we compare the first-order
approximation above to simulation results.

\section{Comparison with simulation results}\label{simul}

In this section we will present a comparison between the
approximations given previously and numerical simulations.
We first describe the Gillespie algorithm used to simulate
  the problem.  We then present and comment on our simulation results, and
  their difference and similarity to the approximate results derived in the previous section.
  
\subsection{Simulations using the Gillespie algorithm}\label{sec:Gillespie}

We describe here, in pseudocode, the algorithm used to simulate a
many-particle system. Steps~(\ref{list:loopMove}) to~(\ref{list:endOfMu})
is a description of the \emph{trial-and-error}-method, outlined in
Ref.~\cite{ambjornsson2008-1} (except that we here use a fixed time step between attempted jumps, see Step~(\ref{list:waittime}) below).

\begin{enumerate}
\item We start by assigning jump rates for each of the $n_b$ directions and
 $n$ crowding particles, and construct a partial sum over them, such that:
 \begin{equation}
   % p_0=0 \quad \text{and} \quad p_a= r_T+(a-1) r_C,\quad a=1,\ldots,n_b \cdot n
   p_0=0, \quad p_a= \sum_{i=1}^a r_i,\quad a=1,\ldots,n_b\cdot n;\quad
   r_i=\begin{cases}
   r_T, & i \leq n_b.\\
   r_C, & i > n_b
 \end{cases}
 \end{equation}
 where $r_T$ and $r_C$ correspond to the jump rate of the tracer and
 crowder particle, respectively.
\item\label{list:loopstart} Place the tracer particle at the center of
 the lattice, and place the remaining particles randomly on the
 lattice, by using the Bebbington algorithm:
 \begin{quote}
   Cycle through all lattice sites (except the position of the
   tracer particle), and place a particle at the current site with
   the probability $a/m$, where $a\in[0,n]$ is the number of particles left
   to place, and $m\in[0,N-1]$ is the number of lattice sites left to cycle
   through~\cite{bebbington1975}.
 \end{quote}
 Set the time equal to zero.
\item\label{list:loopMove} Draw a random number $0<r<1$ with uniform
 distribution, and find the element with label $k$ such that
 \begin{equation}\label{eq:chooseMu}
   p_k \leq rp_{n_b n} < p_{k +1}.
 \end{equation}

\item Convert $k$ to the corresponding particle and direction, and move it,
 provided the chosen site is unoccupied.
\item\label{list:waittime} Update the time to $t \rightarrow t + \frac{1}{p_{n_b n}}$. %, where  $0 < r < 1$ with uniform probability.
\item\label{list:endOfMu} Return to step~(\ref{list:loopMove}) until
 $t \ge t_{\textrm{max}}$.
\item Return to step~(\ref{list:loopstart}) to start the simulation of
 the next trajectory.
\end{enumerate}
As a stop time, we choose $t_{\textrm{max}}=10t_{\min} $ in all
simulations, where $t_{\min}$ is given in \eref{eq:t_min}. The number
of trajectories is denoted by $M$.

Our simulation software package (written in C++) is available from
\url{https://github.com/impaktor/diffusion} and is released under the GNU General Public License (GPL)
\cite{gpl3}.

Based on the tracer particle trajectories, diffusion constants were obtained using
  weighted least-squares fitting to the MSD, see \ref{sec:fitting} for details.

\begin{figure}  \begin{center}
\includegraphics[width=0.7\textwidth]{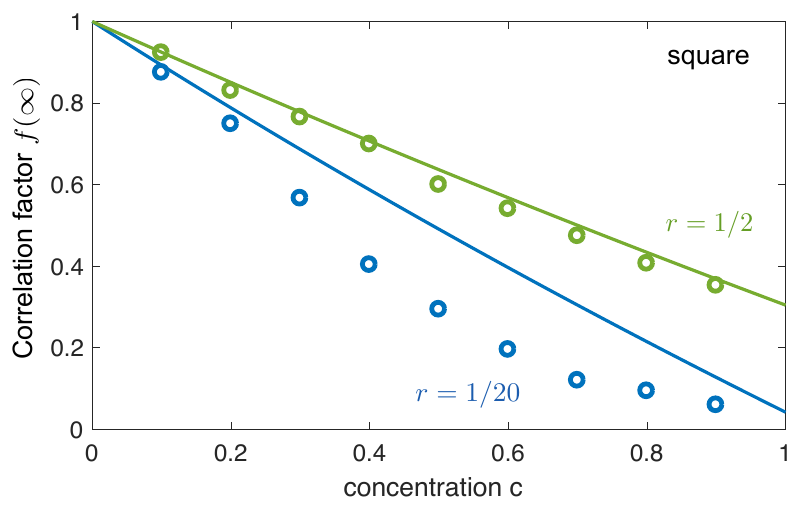}
\caption{Simulations and theoretical estimation for the correlation factor in
  the long time limit for two-dimensional square lattice: The correlation factor, $f$, is shown as a function of the
concentration $c$ for a relative jump rate $r=1/20$ in blue (lower curves) and
$r=1/2$ in green (upper curves). The marks correspond to the Gillespie
simulations ($M=400$ for $r=1/20$ and $M=1000$ for $r=1/2$), while the full curves are determined by the first order approximation given in \eref{fmamm}.}  \label{square}
\end{center}
\end{figure}

\subsection{Comparison between simulations and analytic predictions }

\begin{figure}  \begin{center}
\includegraphics[width=0.7\textwidth]{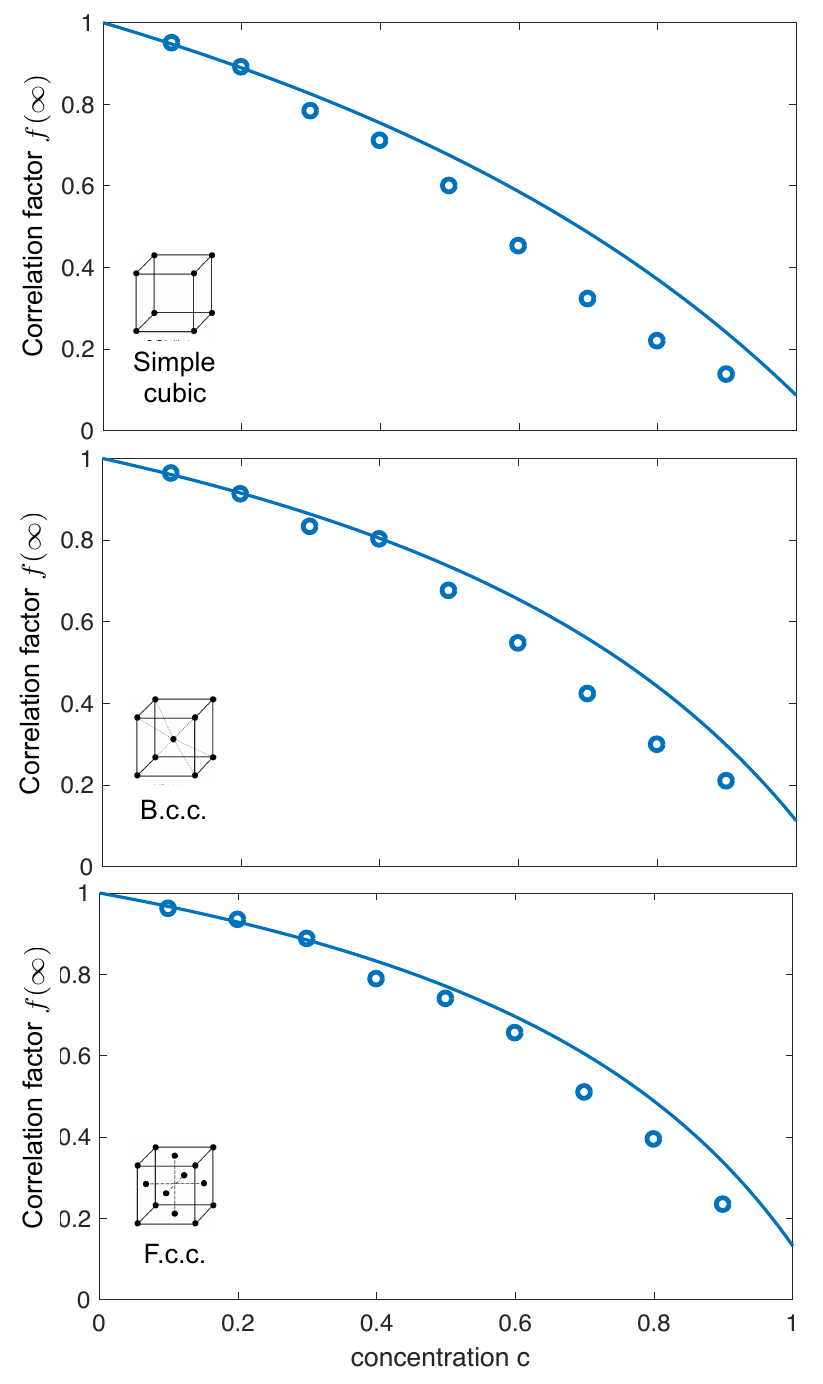}
\caption{Simulations and theoretical predictions for the correlation factor in
  the long-time limit for three-dimensional lattices. Shown is the correlation factor $f$ as a function of the
concentration $c$ for a relative jump rate $r=1/20$. The lattice geometry is
indicated in inset (simple cubic, b.c.c. and f.c.c. lattices). The marks correspond
to the Gillespie simulations ($M=300$ trajectories),while the full curves are determined by the first-order approximation given in \eref{fmamm}.}   \label{cubic}
\end{center}
\end{figure}

For a set of two- and three-dimensional lattices, the correlation
factor is shown in \fref{square} and \fref{cubic}, together with
the corresponding first-order approximation $f_{\text{first-order}}$,
\eref{fmamm}.  We consider only long-time results for the correlation
factor, and study its concentration dependence.
For the two-dimensional square lattice we consider two jump rates, $r=1/20$ and
$r=1/2$.
The lattice size is $400\times 400$ sites with periodic boundary
conditions, and the results are averaged over $400$ trajectories for
$r=1/20$ and $1000$ trajectories for $r=1/2$ (\fref{square}).
For the three-dimensional lattice we set the relative jump rates to
$r=1/20$.
Here, we choose a lattice size of $200\times 200\times 200$ with
periodic boundary conditions, and the results are averaged over an ensemble
of $300$ trajectories (\fref{cubic}).

For all of the lattices considered, we see that the simulation results
are below the first-order approximation $f_{\text{first-order}}$.
Thus, the first-order result \eref{pexpansion} appears to set
an upper bound for the correlation factor. This finding will be useful
in future attempts to calculate exact, or approximate, expressions for
the higher order terms in \eref{dcdcd}. In \fref{square} we see that
upon decreasing $r$ the discrepancy between the simulation and
\eref{fmamm} increases. This indicates non-negligible contributions of
higher orders of the expansion \eref{pexpansion} for slowly diffusing
tracer particles. However, we see that $f_{\text{first-order}}$, in
general, still gives a very good estimate of the correlation factor. 

\section{Summary}

To summarize, we have here presented a general derivation of the PDF,
and MSD, for the position of a tracer particle
diffusing in a crowded system, where the tracer particle has a different
diffusion constant from the crowder particles. The general derivation
is based on a polaron-like mapping like in Ref. \cite{nakazato1980}.
% We showed that the mean field approximation is independent of the
% geometry of the system (network) and only rely on the concentration
% of the crowders.
We showed that the tracer particle PDF can be expressed as an
expansion in terms of crowder particle correlation functions. Focusing
on regular lattices, we then showed that the first order of the
expansion can be evaluated for regular lattices extending the
results of Nakazato and Kitahara \cite{nakazato1980}. We provide
extensive numerical analysis for the different lattices considered,
suggesting that $f_{\text{first-order}}$ overestimates the true
correlation factor.

The present work, besides proposing a generic approach to diffusion in
a crowded network, improves our understanding of diffusion in crowded
regular lattices, especially in a context of slowly diffusive
environments. The potential range of application of this result
extends from biophysics to solid state physics or other types of
dynamics on networks.

\section*{Acknowledgments}

This work was supported by the Swedish Research Council (grant
no. 2009--2924 and 2014--4305), the Crafoord Foundation (grant 20110588), the John Templeton Foundation (grant ID 43467) and French ANR (grant ACHN - C-FLigHT). SP would like to thank Mauro Paternostro and Gabriele De Chiara for their support. GM would like to thank the Department of Astronomy and Theoretical Physics, Lund University, for hospitality.\\

\appendix

\section{Phase rotation of the initial states}\label{sec:phase_rot}

In this appendix we, for completeness, provide a derivation of the result
given in Ref.~\cite{nakazato1980} for the phase-rotated initial states.

Consider \eref{eq:sqrt_x_eq}, which we rewrite according to:
\begin{align}
  G(\mathbf{r},t|\mathbf{r}_0;x)&=\langle \{\emptyset\}\vert \langle
  T_\mathbf{r}\vert W^{-}(x) e^{-\theta S} e^{t\mathcal{L}}
  e^{+\theta S}W^{+}(x)\vert T_{\mathbf{r}_0}\rangle \vert  \{\emptyset\} \rangle\nonumber\\
  &=\langle T_\mathbf{r}\langle \vert \{\emptyset\}\vert W^{-}(x)
  e^{t{\check{\mathcal{L}}}(\theta)} W^{+}(x)\vert
  T_{\mathbf{r}_0}\rangle \vert \{\emptyset\} \rangle,\label{eq:G_rot}
\end{align}
where $\vert T_{\mathbf{r}_0}\rangle \vert \{\emptyset\} \rangle$
denotes the state of no occupancy at all sites except at the tracer
particle position $\mathbf{r}_0$ as before. Also,
\begin{align}
  W^{+}(x)&= e^{-\theta S}\prod_{j=1}^{N} (1+\sqrt{x}\vert C_{\mathbf{u}_j}\rangle\langle 0_{\mathbf{u}_j}\vert )\:,\\
  W^{-}(x)&=\prod_{j=1}^{N} (1+\sqrt{x}\vert
  0_{\mathbf{u}_j}\rangle\langle C_{\mathbf{u}_j}\vert )e^{+\theta S},
\end{align}
and
\begin{equation}
  {\check{\mathcal{L}}}(\theta)=e^{-\theta S}
  \mathcal{L} e^{+\theta S}.
\end{equation}
In the last step in \eref{eq:G_rot} we used the following result
\begin{align}
  e^{-\theta S} e^{t\mathcal{L}} e^{+\theta S}&=e^{-\theta S} [1+\mathcal{L}t+\frac{1}{2!}(\mathcal{L}t)^2+ \cdots ]e^{+\theta S}\nonumber\\
  &=1+t e^{-\theta S}\mathcal{L}e^{+\theta S}+ \frac{t^2}{2!}e^{-\theta S}\mathcal{L}\mathcal{L}e^{+\theta S}+\cdots\nonumber\\
  &=1+t e^{-\theta S}\mathcal{L}e^{+\theta S}+
  \frac{t^2}{2!}e^{-\theta S}{\cal
    L}e^{+\theta S}e^{-\theta S}\mathcal{L}e^{+\theta S}+\cdots\nonumber\\
  & = \exp[te^{-\theta S}\mathcal{L}e^{+\theta
    S}]=\exp[{t\check{\mathcal{L}}}(\theta)],
\end{align}
where $S$ should not depend on the location of the tracer particle
site.  Following Nakazato and Kitahara \cite{nakazato1980} one chooses
\begin{equation}
  S=\sum_{j=1}^{N}S_j,
\label{eq:S}
\end{equation}
where $S_j = \vert C_{\mathbf{u}_j}\rangle\langle 0_{\mathbf{u}_j}\vert
-\vert 0_{\mathbf{u}_j}\rangle\langle C_{\mathbf{u}_j}\vert $.  Now,
\begin{equation}
  e^{-\theta S_j}= 1-\theta S_j +\frac{\theta^2}{2!}S_j^2
  -\frac{\theta^3}{3!}S_j^3 + \frac{\theta^4}{4!}S_j^4 +\cdots
\label{eq:exp_S}
\end{equation}
We have
\begin{align}
  S_j^2&= (\vert C_{\mathbf{u}_j}\rangle\langle 0_{\mathbf{u}_j}\vert
  -\vert 0_{\mathbf{u}_j}\rangle\langle C_{\mathbf{u}_j}\vert )(\vert
  C_{\mathbf{u}_j}\rangle\langle
  0_{\mathbf{u}_j}\vert -\vert 0_{\mathbf{u}_j}\rangle\langle C_{\mathbf{u}_j}\vert )\nonumber\\
  &=-\vert C_{\mathbf{u}_j}\rangle\langle C_{\mathbf{u}_j}\vert -\vert
  0_{\mathbf{u}_j}\rangle\langle 0_{\mathbf{u}_j}\vert =-I_j,
\end{align}
where $I_j =\vert C_{\mathbf{u}_j}\rangle\langle C_{\mathbf{u}_j}\vert
+\vert 0_{\mathbf{u}_j}\rangle\langle 0_{\mathbf{u}_j}\vert $ is an
identity operator which leaves any no-tracer state unchanged. Therefore
\begin{align}
  S_j^3&=-S_j,\nonumber\\
  S_j^4&=(-I_j)(-I_j)= I_j,
\end{align}
etc. \eref{eq:exp_S} now becomes
\begin{align}
  e^{-\theta S_j}&= I_j-\theta S_j -\frac{\theta^2}{2!}I_j
  -\frac{\theta^3}{3!}S_j - \frac{\theta^4}{4!}I_j +
  \cdots  \nonumber\\
  &=\cos\theta \,I_j -\sin\theta\,S_j.
\end{align}
We now have
\begin{align}
  e^{-\theta
    S_j}(1+\sqrt{x}\vert C_{\mathbf{u}_j}\rangle\langle 0_{\mathbf{u}_j}\vert
  )\vert T_{\mathbf{r}_0}\rangle\vert  \{\emptyset\} \rangle &=  [\cos\theta \,I_{\mathbf{u}_j} -\sin\theta\,(\vert
  C_{\mathbf{u}_j}\rangle\langle 0_{\mathbf{u}_j}\vert -\vert
  0_{\mathbf{u}_j}\rangle\langle
  C_{\mathbf{u}_j}\vert )]\nonumber\\
  &\quad \times (1+\sqrt{x}\vert C_{\mathbf{u}_j}\rangle\langle
  0_{\mathbf{u}_j}\vert ) ] \vert T_{\mathbf{r}_0}\rangle \vert  \{\emptyset\}
  \rangle\nonumber\\
  &=[(\cos\theta + \sin\theta \sqrt{x})\nonumber\\
  &\quad+ (\cos\theta \sqrt{x}-\sin\theta)(\vert
  C_{\mathbf{u}_j}\rangle\langle 0_{\mathbf{u}_j}\vert ] \vert
  T_{\mathbf{r}_0}\rangle \vert \{\emptyset\} \rangle,
\end{align}
where we used the fact that $\vert 0_{\mathbf{u}_j}\rangle\langle
C_{\mathbf{u}_j}\vert \vert T_{\mathbf{r}_0},\{\emptyset\} \rangle=0$.
If we now choose $\theta$ such that $\cos \theta \sqrt{x}=
\sin\theta$, i.e., $\tan\theta=\sqrt{x}$, so that $\sin\theta =
\sqrt{x}/\sqrt{1+x}$ and $\cos\theta = 1/\sqrt{1+x}$, then
\begin{align}
  W^{+}(\theta,x)\vert T_{\mathbf{r}_0}\rangle\vert
  \{\emptyset\} \rangle
  &=\prod_{l=1}^{N}(\cos\theta+\sin\theta\sqrt{x})\vert
  T_{\mathbf{r}_0}\rangle
  \vert \{\emptyset\} \rangle\nonumber\\
  &=(1+x)^{(N-1)/2}\vert T_{\mathbf{r}_0}\rangle \vert
  \{\emptyset\} \rangle.
\end{align}
Similarly, one finds,
\begin{equation}
  \langle \{\emptyset\} \vert \langle T_\mathbf{r}\vert  W^{-}(x) =
  \langle \{\emptyset\}\vert \langle T_\mathbf{r}\vert (1+x)^{(N-1)/2}.
\end{equation}
Therefore, \eref{eq:G_rot} becomes:
\begin{equation}
  G(\mathbf{r},t\vert \mathbf{r}_o;x)=(1+x)^{(N-1)} \langle \{\emptyset\}\vert
  \langle T_\mathbf{r} \vert e^{t{\check{\mathcal{L}}(\theta)}}
  \vert T_{\mathbf{r}_0}\rangle \vert \{\emptyset\} \rangle,
\end{equation}
which, when combined with \eref{g}, leads to \eref{eq:rot_L} in the
main text.

\section{Phase-rotated Liouvillian}\label{sec:phase_rot_L}

Consider the phase-rotated Liouvillian given in
\eref{eq:rot_L}. Application of The Baker-Campbell-Hausdorff formula
\cite{Wilcox1967} to this expression gives
\begin{align}
{\check{\mathcal{L}}}&=e^{-\theta S} \mathcal{L} e^{+\theta S}
=\mathcal{L}
-\theta[S,\mathcal{L}]+\frac{\theta^2}{2!}[S,[S,\mathcal{L}]] -\frac{\theta^3}{3!}[S,[S,[S,\mathcal{L}]]]+\cdots\label{eq:BCH}
,
\end{align}
where $\mathcal{L}= r_T\mathcal{U}_T + r_C\mathcal{U}_C$ is defined through
\eref{l} and $S$ is given in \eref{eq:S}.
Consider now the second term on the right-hand side of \eref{eq:BCH}.
After some algebra, one finds
\begin{equation}
[S,\mathcal{L}]= - r_T\mathcal{T},
\end{equation}
where $\mathcal{T}=\mathcal{C} + \mathcal{C}^\dagger$,
and where $\mathcal{C}$ and $\mathcal{C}^\dagger$ are given in \erefs{t},
and (\ref{t+}) in the main text, respectively. Proceeding to higher orders, we find after some lengthier algebra, that
\begin{equation}
[S,[S,\mathcal{L}]]=-2r_T\mathcal{U}_T +2 r_T\mathcal{R},
\end{equation}
where $\mathcal{R}$ is given by \eref{r} in the main text.
The next order term becomes:
\begin{equation}
[S,[S,[S,\mathcal{L}]]]=4 r_T\mathcal{T}.
\end{equation}
Proceeding in this manner, one then gets
\begin{align}
{\check{\mathcal{L}}}&=r_C\mathcal{U}_C +r_T\bigg(\mathcal{U}_T +\theta \mathcal{T} +\frac{\theta
  ^2}{2!}2[-\mathcal{U}_T +\mathcal{R}]-\frac{\theta ^3}{3!}[4 \mathcal{T}]\nonumber\\
&-\frac{\theta
  ^4}{4!}8[-\mathcal{U}_T + \mathcal{R}]+\frac{\theta ^5}{5!}[16 \mathcal{T}]+\ldots \bigg) \nonumber\\
&=r_C\mathcal{U}_C +r_T \bigg(\mathcal{U}_T +\frac{1}{2}[-\frac{(2\theta)^2}{2!}
+\frac{(2\theta)^4}{4!}+\ldots \,][\mathcal{U}_T - \mathcal{R}]\nonumber\\
&+\frac{ \mathcal{T}}{2}[(2\theta)-\frac{(2\theta)^3}{3!}
+\frac{(2\theta)^5}{5!}- \ldots \,]\bigg) \nonumber\\
&=r_C\mathcal{U}_C + r_T\bigg( \mathcal{U}_T +\frac{\cos 2\theta
  -1}{2}[\mathcal{U}_T - \mathcal{R}]\bigg) +r_T \frac{ \sin 2\theta }{2}\mathcal{T} \nonumber\\
&=r_C\mathcal{U}_C+r_T\bigg(\frac{1+\cos 2\theta}{2}\,\mathcal{U}_T +\frac{1-\cos 2\theta}{2}\, \mathcal{R}+\frac{\sin2\theta}{2}\, \mathcal{T}\bigg)\nonumber\\
&=r\mathcal{U}_C+r_T\cos ^2\theta\,\mathcal{U}_T +r_T\sin^2\theta\,
 \mathcal{R}+ r_T\sin\theta\cos\theta\,  \mathcal{T}.
\end{align}
which is the result given in the main text, \eref{ltilde}.

\section{Two-body correlation function}\label{aa}

In this appendix we derive an explicit expression for $\tilde{S}_{1}(\mathbf{q},s)$,
see \eref{s1}, which quantifies two-body correlation
effects in the phase-rotated domain. The derivation is similar to
Ref. \cite{Suzuki2002} where related quantities are
derived for honeycomb and diamond lattices.

In order to take into account two-body correlations we need to explicitly
determine \eref{s1}, which we write:

\begin{eqnarray}
\tilde{\hat{S}}_{1}(\mathbf{q},s)&=&[\tilde{\hat{S}}_{0}(\mathbf{q},s)]^{2}\frac{1}{N}\sum_{\mathbf{r},\mathbf{r}_{0}}e^{-i\mathbf{q\cdot}(\mathbf{r}-\mathbf{r}_{0})} \nonumber\\
&&\times \sum_{\mathbf{b},\mathbf{b'}} [ \tilde{Q}(\mathbf{r},\mathbf{b};s|\mathbf{r}_0,\mathbf{b'}) - \tilde{Q}(\mathbf{r}+\mathbf{b},-\mathbf{b};s|\mathbf{r}_0,\mathbf{b'}) \nonumber\\
&& - \tilde{Q}(\mathbf{r},\mathbf{b};s|\mathbf{r}_0-\mathbf{b}',\mathbf{b'}) + \tilde{Q}(\mathbf{r}+\mathbf{b},-\mathbf{b};s|\mathbf{r}_0-\mathbf{b}',\mathbf{b'})],\label{eq:fvf}
\end{eqnarray}
where we introduced
\begin{equation}
Q(\mathbf{r},\mathbf{\Delta};t|
\mathbf{r}_0,\mathbf{\Delta}_0)=\langle\{\emptyset\}\vert\langle{T}_{\mathbf{r}}\vert\langle{C}_{\mathbf{r}+\mathbf{\Delta}}\vert
e^{t\check{\mathcal{L}}_{0}}\vert{T}_{\mathbf{r}_{0}}\rangle\vert{C}_{\mathbf{r}_{0}+\mathbf{\Delta}_{0}}\rangle\vert\{\emptyset\}
\rangle,\label{eq:Q}
\end{equation}
which is the probability that that the tracer particle is at position $\mathbf{r}$ with the crowder particle at a separation $\mathbf{\Delta}$ at time $t$. The initial tracer particle postion is $\mathbf{r}_0$ and the initial separation vector is $\mathbf{\Delta}_{0}$. We will below derive an explicit master equation for $Q(\mathbf{r},\mathbf{\Delta};t| \mathbf{r}_0,\mathbf{\Delta}_0)$. 
Fourier transforming ($\hat{\bullet}$) \eref{eq:Q} with respect to the tracer position from its initial position, $\mathbf{r}-\mathbf{r}_{0}$, and 
and Laplace transforming ($\tilde{\bullet}$) with respect to time $t$, we have 
\begin{equation}
\tilde{\hat{Q}}(\mathbf{q},\mathbf{\Delta},s| \mathbf{\Delta}_0)=\int_{0}^{\infty}dte^{-st}\frac{1}{N}\sum_{\mathbf{r},\mathbf{r}_{0}}e^{-i\mathbf{q.}(\mathbf{r}-\mathbf{r}_{0})}Q(\mathbf{r},\mathbf{\Delta};t| \mathbf{r}_0,\mathbf{\Delta}_0).\label{eq:fl}
\end{equation}
 \eref{eq:fvf} can now be written as 
\begin{equation}
\tilde{\hat{S}}_{1}(\mathbf{q},s)=\frac{1}{N}[\tilde{\hat{S}}_{0}(\mathbf{q},s)]^{2}\sum_{\mathbf{b},\mathbf{b}'}\left(1-e^{i\mathbf{q}\cdot
    \mathbf{b}'}\right)\left[\tilde{\hat{Q}}(\mathbf{q},\mathbf{b},s\vert\mathbf{b}')-e^{i\mathbf{q}\cdot
    \mathbf{b}}\tilde{\hat{Q}}(\mathbf{q},-\mathbf{b},s\vert\mathbf{b}')\right].
\end{equation}
To first order in $\gamma^2$, the structure factor is now
\begin{equation}
\tilde{\hat{S}}(\mathbf{q},s)\approx \tilde{\hat{S}}_{0}(\mathbf{q},s)+ \gamma^2\tilde{\hat{S}}_{1}(\mathbf{q},s) + \mathcal{O}[\mathbf{\gamma}^{4}].
\end{equation}
The contribution to the MSD of $\tilde{S}_{1}(\mathbf{q},s)$ is connected
to the second order term in the expansion of $\tilde{\hat{S}}(\mathbf{q},s)$ in $\mathbf{q}$ (c.f. \eref{chat}).
For all lattices considered here, we have that if $\mathbf{b}$ is a nearest neighbour, then $-\mathbf{b}$ is also a nearest neighbour vector. Using this fact, and that $\tilde{S}_{0}(\mathbf{q},s)\sim1/s+\mathcal{O}[\mathbf{q}^{2}]$ the second order expansion of $S_1$ becomes
\begin{equation}
\tilde{\hat{S}}_{1}(\mathbf{q},s)= \frac{1}{s^{2}}G(s)+\mathcal{O}[\mathbf{q}^{4}]
\end{equation}
with 
\begin{equation}
G(s)=\sum_{\mathbf{b},\mathbf{b}'}(\mathbf{q\cdot b})(\mathbf{q 
\cdot b}')\tilde{\hat{Q}}(\mathbf{0},\mathbf{b},s\vert\mathbf{b}').\label{eq:ok-1}
\end{equation}
Thus, in order to determine the MSD we need to know determine the $\mathbf{q}\rightarrow 0$ limit of the quantity $\tilde{\hat{Q}}(\mathbf{q},\mathbf{\Delta},s| \mathbf{\Delta}_0)$.

Consider now the definition \eref{eq:Q}, and let us derive a master equation for $Q(\mathbf{r},\mathbf{\Delta};t| \mathbf{r}_0,\mathbf{\Delta}_0)$. To that end, we take the derivative with respect to time and use the operator identity $\partial_t \exp[t \check{\mathcal{L}}] = \check{\mathcal{L}} \exp[t \check{\mathcal{L}}]$. By further using the orthogonality relation, \eref{ortho}, we arrive at 
\begin{multline}
\partial_{t}Q(\mathbf{r},\mathbf{\Delta};t)=r_{T}c\sum_{\mathbf{b}}\delta_{\mathbf{\Delta},\mathbf{b}}\left(Q(\mathbf{r}+\mathbf{\Delta},-\mathbf{\Delta};t)-Q(\mathbf{r},\mathbf{\Delta};t)\right)+\sum_{\mathbf{b}}\left(1-\delta_{\mathbf{\Delta},\mathbf{0}}-\delta_{\mathbf{\Delta},\mathbf{b}}\right)\\
\times\bigg[r_{T}(1-c)\left(Q(\mathbf{r}+\mathbf{b},\mathbf{\Delta}-\mathbf{b};t)-Q(\mathbf{r},\mathbf{\Delta};t)\right)+r_{C}\left(Q(\mathbf{r},\mathbf{\Delta}-\mathbf{b};t)-Q(\mathbf{r},\mathbf{\Delta};t)\right)\bigg].\label{eq:two_body_master_eq}
\end{multline}
In the derivation above, we left the initial conditions
  implicit, and used that $Q(\mathbf{r},\mathbf{\Delta}=0;t| \mathbf{r}_0,\mathbf{\Delta}_0)$=0, \cite{Suzuki2002} i.e. two particles cannot occupy the same lattice site. The first term on the right hand side above originates from the $\mathcal{R}$ operator. The $\delta$-functions appearing in the remaining terms make sure there cannot be jumps into and out of the ``forbidden'' state $\mathbf{\Delta}=0$. 

To proceed, we take the Laplace
and Fourier transforms [see \eref{eq:fl}] of \eref{eq:two_body_master_eq}. We also take the Fourier-transform with respect to $\mathbf{\Delta}$ (Fourier variable $\mathbf{p}$). By rearranging the terms in the transformed version of \eref{eq:two_body_master_eq} and by  subsequently performing the inverse Fourier-transform with respect to $\mathbf{p}$ we arrive at
\begin{multline}
\tilde{\hat{Q}}(\mathbf{q},\mathbf{\Delta},s\vert\mathbf{\Delta}')=  e^{-i\mathbf{q}
\cdot \mathbf{r}_{0}}\omega(\mathbf{q},\mathbf{\Delta}-\mathbf{\Delta}',s)+\sum_{\mathbf{b}}\bigg[r_{T}c\left(\omega(\mathbf{q},\mathbf{\Delta}+\mathbf{b},s)e^{-i\mathbf{q}\cdot \mathbf{b}}-\omega(\mathbf{q},\mathbf{\Delta}-\mathbf{b},s)\right) \\
  +r_{T}(1-c)\left(\omega(\mathbf{q},\mathbf{\Delta}-\mathbf{b},s)-\omega(\mathbf{q},\mathbf{\Delta},s)e^{-i\mathbf{q}\cdot \mathbf{b}}\right)+r_{C}\left(\omega(\mathbf{q},\mathbf{\Delta}-\mathbf{b},s)-\omega(\mathbf{q},\mathbf{\Delta},s)\right)\bigg]\\
  \times\tilde{\hat{Q}}(\mathbf{q},\mathbf{b},s\vert\mathbf{\Delta}')\label{eq:gt}
\end{multline}
with 
\begin{equation}
\omega(\mathbf{q},\mathbf{\Delta},s)=\frac{1}{(2\pi)^{d}} \int d^d
p \frac{e^{i\mathbf{p}\cdot\mathbf{\Delta}}}{s-r_{T}(1-c)\sum_{\mathbf{b}}\left(e^{-i\mathbf{q}\cdot \mathbf{b}}e^{i\mathbf{p}\cdot \mathbf{b}}-1\right)-r_{C}\sum_{\mathbf{b}}\left(e^{-i\mathbf{p}\cdot \mathbf{b}}-1\right)}.
\end{equation}
where $\int d^dp=(l_1\cdots l_d) \int_{-\pi/l_1}^{\pi/l_1}dp_1\cdots \int_{-\pi/l_d}^{\pi/l_d}dp_d$ with $l_i = a v_i$ ($i=1,\ldots,d$) and $\mathbf{v}$ is given in \tref{lambda}. By letting $\mathbf{q}\rightarrow 0$ in \eref{eq:gt}, we have an equation for the sought quantity, $\tilde{\hat{Q}}(\mathbf{q}=0,\mathbf{b},s\vert\mathbf{b}')$, see \eref{eq:ok-1}. More precisely, by letting $\mathbf{q}\rightarrow 0$ and by setting $\mathbf{\Delta'} = \mathbf{b'}$ and $\mathbf{\Delta} = \mathbf{b''}$ in \eref{eq:gt} for all nearest neighbour vectors $\mathbf{b''}$, we get (for each $\mathbf{b'}$) a linear system of equations. The size of this system of equations is equal to the number of nearest neighbours for each lattice, in general. 

For all lattices considered herein, we are fortunate that one need not solve explicitly the full set of systems of equations provided by \eref{eq:gt}. Instead, one will get a scalar equation for the quantity $G(s)$, see \eref{eq:ok-1}, directly. To show this, we let $\mathbf{q}\rightarrow 0$ in \eref{eq:gt} and set $\mathbf{\Delta}=\mathbf{b}$ and $\mathbf{\Delta}'=\mathbf{b}'$. By subsequently multiplying \eref{eq:gt} by $(\mathbf{q\cdot b})(\mathbf{q 
\cdot b}')$ and summing over $\mathbf{b}$ and $\mathbf{b}'$ we find that
\begin{multline}
G(s)=\sum_{\mathbf{b},\mathbf{b}'}(\mathbf{q}\cdot \mathbf{b})(\mathbf{q}\cdot \mathbf{b}')R(\mathbf{b},\mathbf{b}',s)\\ +\left( -2r_Tc + \left[r_{T}(1-c)+r_{C}\right]\right) 
\sum_{\mathbf{b},\mathbf{b}',\mathbf{b}''}(\mathbf{q}\cdot \mathbf{b})(\mathbf{q}\cdot \mathbf{b}'')R(\mathbf{b},\mathbf{b}'',s)\tilde{\hat{Q}}(\mathbf{0},\mathbf{b},s\vert\mathbf{b}'),\label{eq:gt-intermediate}
\end{multline}
where 
\begin{equation}
R(\mathbf{b},\mathbf{b}',s)=\frac{1}{(2\pi)^{d}} \int d^d p \frac{\sin\left(\mathbf{p}\cdot \mathbf{b}\right)\sin\left(\mathbf{p}\cdot \mathbf{b}'\right)}{s-\left[r_{T}(1-c)+r_{C}\right] \sum_{\mathbf{\tilde{b}}}\left(e^{i\mathbf{p}\cdot \mathbf{\tilde{b}}}-1\right)}.\label{eq:eq_for_R}
\end{equation}
In the derivation leading up to \eqref{eq:gt-intermediate} we used the facts that 
\begin{equation}
\frac{1}{(2\pi)^{d}}\int d^d p \frac{\sin\left(\mathbf{p}\cdot \mathbf{b}\right)\cos\left(\mathbf{p}\cdot \mathbf{b}'\right)}{s-\left[r_{T}(1-c)+r_{C}\right] \sum_{\mathbf{\tilde{b}}}\left(e^{i\mathbf{p}\cdot \mathbf{\tilde{b}}}-1\right)} = 0
\end{equation}
and that 
\begin{equation}
\frac{1}{(2\pi)^{d}}\int d^d p \frac{\sin\left(\mathbf{p}\cdot \mathbf{b}\right)}{s-\left[r_{T}(1-c)+r_{C}\right] \sum_{\mathbf{\tilde{b}}}\left(e^{i\mathbf{p}\cdot \mathbf{\tilde{b}}}-1\right)} = 0
\end{equation}
for the lattices considered herein. The
expression above follows, for instance, by explicit calculations for each of
the lattices using the trigonometric identities~\cite{abramowitz1964}
\begin{eqnarray}\label{eq:trig1}
\sin(z_1+z_2) &=& \sin(z_1)\cos(z_2) + \cos(z_1)\sin(z_2)\nonumber \:, \\
\cos(z_1+z_2) &=& \cos(z_1)\cos(z_2) - \sin(z_1)\sin(z_2) \:,
\end{eqnarray}
and (using the expressions above)
\begin{eqnarray}\label{eq:trig2}
\sin(z_1+z_2+z_3) &=& -\sin(z_1)\sin(z_2)\sin(z_3) + \sin(z_1)\cos(z_2)\cos(z_3) \nonumber\\
&&+ \cos(z_1) \sin(z_2)\cos(z_3) + \cos(z_1) \cos(z_2)\sin(z_3) ,
\end{eqnarray}
\begin{eqnarray}\label{eq:trig3}
\cos(z_1+z_2+z_3) &=& \cos(z_1)\cos(z_2)\cos(z_3) - \cos(z_1)\sin(z_2)\sin(z_3)\nonumber\\
&& - \sin(z_1) \sin(z_2)\cos(z_3) - \sin(z_1) \cos(z_2)\sin(z_3) ,
\end{eqnarray}
as well as the fact that ($v=x,y,z$ and $\alpha,\beta$ are arbitrary real numbers)  
\begin{equation}\label{eq:trig4}
\int_{-\pi/l_v}^{\pi/l_v} \frac{{dp_v}}{2\pi} \frac{ \sin\left( \alpha p_v \right)\cos\left(\beta p_v\right)}{s-\left[r_{T}(1-c)+r_{C}\right]\sum_{\mathbf{b}}\left(e^{i\mathbf{p}\cdot \mathbf{b}}-1\right)}=0,
\end{equation}
and 
\begin{equation}\label{eq:trig5}
\int_{-\pi/l_v}^{\pi/l_v} \frac{{dp_v}}{2\pi} \frac{ \sin\left( \alpha p_v
  \right)}{s-\left[r_{T}(1-c)+r_{C}\right]\sum_{\mathbf{b}}\left(e^{i\mathbf{p}\cdot
      \mathbf{b}}-1\right)}=0.
\end{equation}
These results follow, since the denominators above are invariant under
$p_x\rightarrow -p_x$, $p_y\rightarrow -p_y$ and $p_z\rightarrow -p_z$ for all
lattices herein. 

To proceed we write $\mathbf{b}=\pm a\mathbf{v}$ (c.f. \tref{tabla}), in \eref{eq:eq_for_R} and \eref{eq:gt-intermediate}. We also use the fact that for all lattices considered we have  
\begin{equation}
\sum_{\mathbf{v}}(\mathbf{q}\cdot\mathbf{v})R(a\mathbf{v}',a\mathbf{v},s)=(\mathbf{q}\cdot \mathbf{v}')V(s) \label{eq:magic_eq}
\end{equation}
where the quantity $V(s)$ depends on the lattice (see below). The expressions for $V(s)$ are calculated explicitly for each lattice by using \eref{eq:trig1} to \eref{eq:trig4}. We then obtain the results for $V(s)$ as given in the main text, after a change of integration variables $p_x\rightarrow p_x/l_x$, $p_y\rightarrow p_y/l_y$ and $p_x\rightarrow p_z/l_z$. 
By combining \eref{eq:magic_eq} and \eref{eq:gt-intermediate}, we  get a closed form equation for $G(s)$: 
\begin{equation}
G(s)=4a^2 V(s) \sum_{\mathbf{v}}(\mathbf{q}\cdot \mathbf{v})^{2}+2\left[r_{T}(1-3c)+r_{C}\right]V(s)G(s),\label{eq:gt-2}
\end{equation}
Finally, solving the equation above, and using the explicit nearest neighbour
vectors listed in \tref{lambda}, we arrive at: 
\begin{equation}
G(s)=\frac{ 2 a^2 n_b V(s) \vert\mathbf{q}\vert^{2}/d}{1-2\left[r_{T}(1-3c)+r_{C}\right]V(s)}.\label{eq:gt-2-1}
\end{equation}
where $\vert\mathbf{q}\vert^{2}= q_x^2+q^2_y+q_z^2$ for 3D lattices, $\vert\mathbf{q}\vert^{2}= q_x^2+q^2_y$ for 2D lattices and $\vert\mathbf{q}\vert^{2}= q_x^2$ for 1D lattices. Above, $n_b$ and $d$ are the number of neighbouring sites and the lattice dimension, respectively.

\section{First order correlation factor: time dependence and long time limit}\label{tt}

We detail here the derivation of the first order results for the
correlation factor, \erefs{flong} and (\ref{fmamm}), respectively, for
a linear (one-dimensional) lattice and for higher-dimensional
lattices.

\emph{One-dimensional lattice:} For a linear lattice the integral
$V(s)$, given in \eref{cub}, can be evaluated:
\begin{align}
  V(s) & = \frac{1}{2\pi} \int_{-\pi}^{\pi} dp \frac{\sin^2\left( p \right)}
  {s-w\lambda(p) } = \frac{1}{ 4 w^2}\left(s+2w-\sqrt{s (s+4 w)}\right) \nonumber\\
  & = \frac{1}{s+2w+\sqrt{s (s+4 w)} }\:,
\label{sbis}
\end{align}
where $\lambda(p)= 2\left( \cos p -1 \right)$ and $w=r_C + r_T (1-c)$.
\eref{A1} becomes
\begin{equation}
  A_1(t)t =  -  4 n_ba^2 {L}^{-1}\left\{  \frac{1}{s^2}\frac{1}{s+2b+\sqrt{s (s+4w)}}
  \right\},
\label{212156}
\end{equation}
with $b=2r_T c$. Focusing on the long-time limit ($s\to 0$) we find, after 
inversion back to the time-domain, that:
\begin{equation}
  A_1(t) \underset{t\to \infty}{\sim} \frac{A_0}{\gamma^2}
  \left( -1 + \frac{1}{c}\sqrt{\frac{r+1-c}{r_T \pi t}} + O(t^{-1})
  \right),
\label{211}
\end{equation}
where $A_0$ is given in \eref{eq:A0} and $\gamma^2$ is defined in
\eref{eq:gamma2}. Inserting the expression above into \eref{l1d0}
yields \eref{flong} in the main text.

\emph{Two- and three-dimensional lattices:} In the higher-dimensional
lattices ($d\ge2$) the integral $V(s)$, given in \erefs{cub}, (\ref{bcc}) and (\ref{fcc})
cannot be evaluated. However, the long time behavior may still be
derived. Considering the asymptotic behavior of \eref{A1} we have
\begin{equation}
  A_{1}(t)t   \underset{t\to\infty}{\sim}  {L}^{-1}\left\{ \frac{1}{s^2}
    \frac{ -4n_ba^2 V(0) }{1- 2 (r_T (1-3c) + r_C )  V(0)  } \right\},
\label{s12}
\end{equation}
Hence, if we define $\beta$ through $\left(\beta+1\right)^{-1} = 2r_T(r+1-c)  V(0) $, we arrive at
\begin{equation}
  A_1(t) \underset{t\to \infty}{\sim} - \frac{A_0}{\gamma^2}
  \left( \frac{ 2 c }{2c + (r + 1-c )  \beta } \right),
\label{eq:A1_final_2d}
\end{equation}
which yields \eref{fmamm} in the
main text. 
Using \eref{cub}, \eref{bcc} and \eref{fcc} we get the explicit expressions for $\beta$ according to
\begin{align}
(\beta_{\text{hyp}}+1)^{-1} & = \int_{-\pi}^{\pi} \frac{dp_x}{2\pi} \int_{-\pi}^{\pi} \frac{dp_y}{2\pi}\int_{-\pi}^{\pi} \frac{dp_z}{2\pi}
\frac{\sin^{2}(p_{x})}{\sum_{\mathbf{v}}(1-\cos(\mathbf{p}\cdot\mathbf{v}))} \:,\\
(\beta_{\text{bcc}}+1)^{-1}& =4\int_{-\pi}^{\pi} \frac{dp_x}{2\pi} \int_{-\pi}^{\pi} \frac{dp_y}{2\pi}\int_{-\pi}^{\pi} \frac{dp_z}{2\pi}\frac{\sin^{2}(p_{x}/\sqrt{3})\cos^{2}(p_{y}/\sqrt{3})\cos^{2}(p_{z}/\sqrt{3})}{\sum_{\mathbf{v}}(1-\cos(\mathbf{p}\cdot \mathbf{v}))} \:, \\
(\beta_{\text{fcc}}+1)^{-1}& =2\int_{-\pi}^{\pi} \frac{dp_x}{2\pi} \int_{-\pi}^{\pi} \frac{dp_y}{2\pi}\int_{-\pi}^{\pi} \frac{dp_z}{2\pi}\frac{\sin^{2}(p_{x}/\sqrt{2})\cos(p_{y}/\sqrt{2})\left(\cos(p_{y}/\sqrt{2})+\cos(p_{z}/\sqrt{2})\right)}{\sum_{\mathbf{v}}(1-\cos(\mathbf{p}\cdot \mathbf{v}))} \:.
\end{align}

For square lattices the two-dimensional integral for $\beta_{\text{hyp}}$ can be
evaluated exactly as $\beta_{\text{square}}=2/(\pi-2)$, whereas for
the three-dimensional lattices the integrals above are numerically calculated, see \tref{tabla}.

\section{Fitting procedure for the diffusion constant}\label{sec:fitting}

\begin{figure}
\begin{center}
\includegraphics[width=0.6\textwidth]{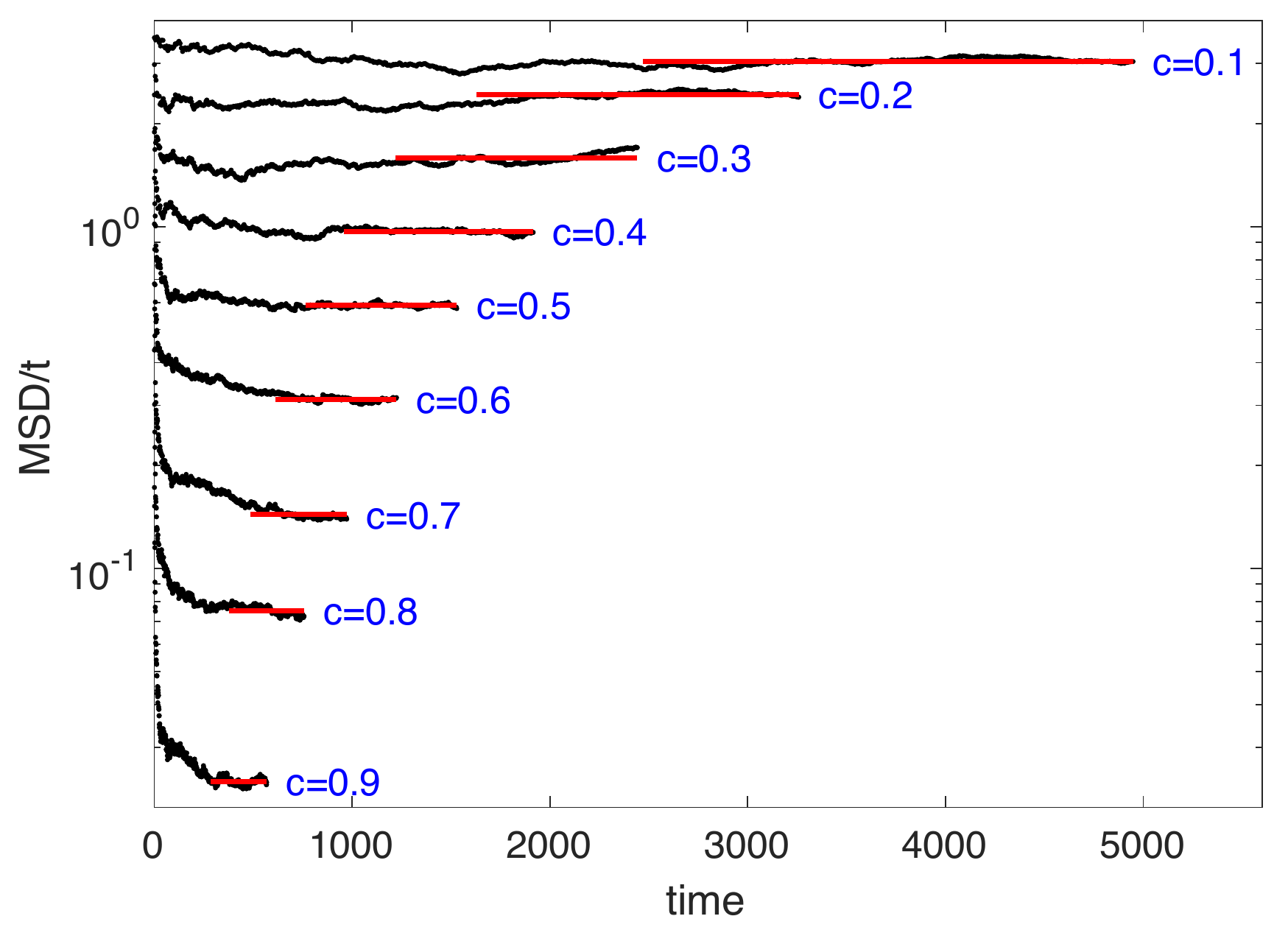}
\caption{Example of MSD fits to simulation data. $\mathcal{MSD}(t)/t$ (logarithmic scale) as a function of time for a square lattice with $r=1/20$ for different concentrations $c$ as indicated. The black marks represent the result from the simulation, whereas the red lines indicate the outcome of the fitting procedure as well as the range of data for which the fitting procedure was applied, see \eref{eq:t_min}.}\label{fit}
\end{center}
\end{figure}
From the set of trajectories generated using the algorithm from
\sref{sec:Gillespie}, we calculate squared displacements $y^{(m)}_i$,
where $m$ ($1\le m\le M$) labels trajectories and $i$ labels sampling
times ($i\in[1,n_s]$, $n_s$ being the number of sampling times). The
estimated MSD then is
\begin{equation}
\bar{y}_i  = \frac{1}{M}\sum_{m=1}^M
 y^{(m)}_i.
\end{equation}
We are here interested in the long-time behavior of the MSD. To
 that end, we keep only data from sampling times for which $t> t_{\min }$ with:
\begin{equation}\label{eq:t_min}
t_{\min} = \frac{10}{c \min(r_C,r_T)}.
\end{equation}

To fit the MSD we used $\chi^2$ regression for uncorrelated data
\cite{Press2007}. We have for the estimated diffusion constant $\hat{D}$:
\begin{equation}
\hat{D} = \frac{1}{2d}\frac{\sum_i \bar{y}_i t_i/\bar{\sigma}_i^2}{\sum_i t_i^2/\bar{\sigma}_i^2}
\end{equation}
with
\begin{equation}
\bar{\sigma}_i^2 = \frac{1}{M(M-1)}\sum_{m=1}^M (y_i^{(m)}-\bar{y}_i)^2\,,
\end{equation}
the variance of $\bar{y}_i$. For all fits we used
%
%BS changed:
%$n_s/M= 10^{-1}$
$n_s = 10^{-1}M$
equidistant sampling times. In \fref{fit} we show a few examples of
fits, here for the data corresponding to \fref{square} (upper panel,
blue marks).

\section*{References}
\bibliographystyle{iopart-num}
\bibliography{references}

\end{document}